\documentclass[pra,twocolumn]{revtex4-1} 
\usepackage{amsmath}  % needed for \tfrac, \bmatrix, etc.
\usepackage{amsfonts} % needed for bold Greek, Fraktur, and blackboard bold
\usepackage{graphicx} % needed for figures

\newcommand{\newvec}[1]{\boldsymbol{#1}}

\begin{document}

% Be sure to use the \title, \author, \affiliation, and \abstract macros
% to format your title page.  Don't use lower-level macros to  manually
% adjust the fonts and centering.

\title{Singularities and internal rotational dynamics of electron beams}

% In a long title you can use \\ to force a line break at a certain location.

 \author{ 
  D. Velasco-Mart\'inez$^1$,
  V. G. Ibarra-Sierra$^2$,
  J. C. Sandoval-Santana$^2$,
  J.L. Cardoso$^1$,
  and
  A. Kunold$^1$
 }

 \affiliation{ 
  $^1$ \'Area de F\'isica Te\'orica y Materia Condensada,
   Universidad Aut\'onoma Metropolitana  Azcapotzalco,
   Av. San Pablo 180, Col. Reynosa-Tamaulipas, Azcapotzalco,
   02200 Cuidad de M\'exico, M\'exico 
   \\
   $^2$ Departamento de F\'isica, Universidad Aut\'onoma Metropolitana
   Iztapalapa, Av. San Rafael Atlixco 186, Col. Vicentina, Iztapalapa,
   09340 Cuidad de M\'exico, M\'exico
   } 

\date{\today}

\begin{abstract}
We study the internal rotational dynamics of
electronic beams in relation to the phase singularities of their wave functions.
Given their complex singularity  structure,
Hermite-Gaussian beams and other superpositions of Laguerre-Gaussian modes
are studied here.
We show that by inspecting the lowest non-vanishing terms of the wave function
near the singularity it is possible to infer the structure of the Bohmian streamlines.
Conversely, starting from a map of the electron's Bohmian velocities, we demonstrate
that it is possible to derive the form of the electron's wave function near the singularity.
We outline a procedure that could yield an experimental method to determine
the main parameters of the electron's wave function close to a singularity.
\end{abstract}

\maketitle

\section{Introduction} 

The charged particle subject to a uniform and static
magnetic field\cite{fock1928,PSP:1733488,landau1977} is at the heart
of many  physical theories. 
The quantum Hall \cite{Klitzing1980} and related
effects such as De Haas-Van Alphen effect \cite{Holstein1973},
Shubnikov-De Haas oscillations \cite{Fang1968} and the
fractional quantum Hall effect \cite{PhysRevLett.50.1395} 
can not be understood without it.
However, solid state systems are too complex to provide
precise information of the internal dynamics of the Landau levels.
Inspired by optical-vortex-beams, 
electron-vortex-beams
have recently gained great attention \cite{PhysRevLett.99.190404,McMorran192,
ISI:000276205000040, ISI:000341052900013,grillo2015, ISI:000281824900033}
not only because they
are expected to provide new capabilities to electron microscopes but
also for allowing the observation of the electrons' isolated quantum states.

Whereas electron vortex beams may be understood as  solutions
of the Schr\"odinger equation, their optical counterparts are
paraxial solutions of the Helmholtz equation
\cite{McMorran192,ISI:000276205000040,ISI:000341052900013}.
In contrast to solid state systems, Laguerre-Gaussian (LG) beams allow
for more precise and accurate
observation of the internal dynamics of the Landau levels because they
are generated in the rather controlled environment of a
transmission electron microscope (TEM).

Among other techniques, electron vortex beams with helical wavefront carrying
a large orbital angular momentum (as LG beams)
are produced passing a standard electron beam through a holographic mask in
TEMs
\cite{PhysRevLett.114.034801,McMorran192,ISI:000276205000040,ISI:000341052900013}.
For example, LG beams are created by
using a diffraction grating
with a dislocation \cite{ISI:000341052900013} that confers electrons
a certain orbital angular momentum (OAM). In order to get actual Fock-Darwin states,
the beam waist is chosen to match their diameter in a region where
the axial magnetic field is uniform.
When LG beams
propagate along the direction of a uniform magnetic field,
they behave as stable solutions of the Schr\"odinger equation.
LG beams correspond in fact to the Fock-Darwin states
that in turn are stationary solutions of the Shr\"odinger equation of a 2D
charged particle in a magnetic field.
Thereby most of their features remain constant
and, as for their optical counterparts, the beam profile
is maintained as electrons propagate. Among these quantities, the
OAM is preserved.

Bessel electron beams \cite{PhysRevX.4.011013},
Airy beams  \cite{ISI:000315312900033} and Hermite-Gaussian (HG)
beams
\cite{PhysRevLett.109.084801}  have been studied and produced
in the laboratory. HG modes, for example, may serve
to probe the OAM of electron beams.  All of these beams can be
understood as linear combinations of LG states carrying different
amounts of OAM.

Phase singularities play a crucial role in the formation of vortices in light.
They have been widely studied since the pioneering work
of Dirac \cite{Dirac60}, Aharonov, Bohm \cite{PhysRev.115.485},
Nye \cite{Nye165} and Berry \cite{Berry45}.
Mathematically, optical singularities are regions in the domain of the 
electromagnetic field
where the phase is indeterminate; as a consequence the field vanishes
in these regions
of space.
Phase singularities are also present in electron beams and manifest
themselves as vortices of density currents that turn around a zero
of the wave function.
Due to their cylindrical symmetry, LG beams can be made to have
large OAM. Their wavefront
spirals around a phase singularity  located at the centre of the
wave front \cite{Schachinger201517,Gbur2003117}.
Singularities may be characterized by their topological charge, an
integer number that counts the number of times the wave front
makes a full revolution in one wavelength.
In the case of LG beams and other eigenstates
of the OAM the topological charge matches the OAM.
Some linear combinations of LG modes are very interesting
in this regard because, not carrying a distinctive angular
momentum, they may present rich and complex singularity arrangements.
Around each singularity, a density current vortex is formed whose  properties
fully depend on the mathematical form of the singularity.

Given the great amount of knowledge on
optical vortices, they where the first candidates for 
OAM-based spectroscopic techniques. However,
optical techniques have been proven to be very inefficient,
owing to the weak optical multipolar transitions
\cite{PhysRevA.83.065801}.
In contrast,  it has been theoretical \cite{PhysRevLett.108.074802}
and experimentally\cite{ISI:000281824900033} demonstrated that the OAM of electron
beams can be transferred to atomic electrons having an observable
effect on the magnetic circular dichroism.
It has also been suggested, that by creating different superpositions of
LG beams \cite{1367-2630-17-9-093015}, it is possible to tailor
the internal electron currents. For instance, by superimposing various LG states it is possible to
produce off-axis density current vortices \cite{PhysRevX.4.011013}.
This would allow addressing specific atoms in the sample by
novel TEM techniques such as chiral-specific electron-vortex-beam
spectroscopy \cite{PhysRevA.88.031801}.
 
In this paper we study the relation between the internal
rotational dynamics of electron beams and their singularities.
We study the features of the canonical and kinetic currents
\cite{1367-2630-17-9-093015,PhysRevLett.113.240404} close
to the phase singularities.
We show that the properties of the electron density currents
are completely determined by the singularities' structure and
topological charge.
To prove our results we have studied HG and related modes.
The type of states that give rise to stable vortex
excitations in Bose-Einstein condensates (BEC)\cite{PhysRevX.4.011013}
are also studied here given their interesting singularity structure. 

Our results show that the mathematical structure of each singularity,
completely determines the vortex structure
in its vicinity.
In particular, we demonstrate
that it is possible to guess the shape of the Bohmian
trajectories close to a singularity by inspecting the wave function's
structure near it.
Conversely, we show that it is possible to obtain the wave
function's mathematical structure through the Bohmian streamlines
around a vortex.
This could lead to an experimental
method to determine the electron's wave function
close to a singularity. 

This paper is organized as follows. In Sec. \ref{fdstates}
we study Fock-Darwin states and introduce the rising and
lowering operators that generate them. These operators are
very practical for obtaining general expressions of the kinetic and canonical
density currents in superpositions of LG modes.
The mathematical form of HG and
related modes as well as BEC states are
introduced in Sec. \ref{compbeams}.
Sec. \ref{singu:struc} presents the main results
regarding the connection of the phase singularities'
structure and the kinetic and canonical density currents.
In Sec. \ref{res:disc} we apply these results to the characterisation of the
the singularities found in HG and BEC modes.
We present the outline of a method to
characterize the mathematical structure of singularities
from the Bohmian streamlines
 in Sec. \ref{outline}.
In Sec. \ref{conclusions} we summarize the results
and propose general conclusions. 

\section{LG modes. Fock-Darwin states}
\label{fdstates}

The starting point is
the Hamiltonian of an electron in a uniform magnetic
field given by 
\begin{equation} \label{ham:one}
\hat{H} = \frac{1}{2m} \left( \hat{\boldsymbol{p}} +e 
\hat{\boldsymbol{A}} \right)^2,
\end{equation}
where the vector potential in the symmetric gauge is given by
\begin{equation}
\hat{\newvec{A}} = - \frac{B}{2} \hat{y} \boldsymbol{i} + \frac{B}{2} \hat{x} \boldsymbol{j} .\label{sym:gauge}
\end{equation}
This particular choice makes the Hamiltonian (\ref{ham:one})
invariant under rotations.
By expanding the momentum and position components,
the Hamiltonian takes the form
\begin{equation} \label{Schr:hamilton}
\hat{H} = \frac{1}{2m}\left( \hat{p}_x^2 +\hat{p}_y^2 \right) + 
\frac{m \omega^2 }{8}\left( \hat{x}^2 +\hat{y}^2 \right) 
+ \frac{\omega}{2}\ \left( \hat{x} \hat{p}_y -\hat{y} \hat{p}_x \right),
\end{equation}
where $\omega = eB /m$ is the cyclotron frequency, the
position and momentum operators follow the standard commutation
relations $\left[\hat x,\hat y\right]=\left[\hat p_x,\hat p_y\right]=0$ and
$\left[\hat x,\hat p_x\right]=\left[\hat y,\hat p_y\right]=i\hbar$.
In the equation above
we immediately identify 
the $z$-component of the angular momentum 
$\hat{L}_{z} = \hbar l_z=\hat{x} \hat{p}_y - \hat{y} \hat{p}_x$.
A bit of algebra  shows that
it commutes with the 
Hamiltonian, i.e.  $\left[ \hat{H}, \hat{L}_{z} \right] = 0$, and therefore  must 
be a conserved quantity.  This is a direct consequence of the adopted  gauge.

To simplify  the Hamiltonian
we define the following rising and lowering operators
\begin{eqnarray}
\hat{b} &=& 
\frac{1}{2\sqrt{2}l_B}\left[\left(\hat x+i\hat y\right)
+i\frac{2}{m\omega}\left(\hat p_x+i\hat p_y\right)\right], \label{op:b} \\
\hat{b}^\dag &=&\frac{1}{2\sqrt{2}l_B}\left[\left(\hat x-i\hat y\right)
-i\frac{2}{m\omega}\left(\hat p_x-i\hat p_y\right)\right], \label{op:bt}\\
\hat{c}&=&
\frac{1}{2\sqrt{2}l_B}\left[\left(\hat x-i\hat y\right)
+i\frac{2}{m\omega}\left(\hat p_x-i\hat p_y\right)\right],\label{op:c}\\
\hat{c}^\dagger &=& 
\frac{1}{2\sqrt{2}l_B}\left[\left(\hat x+i\hat y\right)
-i\frac{2}{m\omega}\left(\hat p_x+i\hat p_y\right)\right],\label{op:ct}
\end{eqnarray}
where $l_B=\sqrt{\hbar/m\omega}$ is the magnetic length.
It can readily be verified that they follow the usual commutation rules 
$\left[\hat b,\hat b^\dagger\right]=\left[\hat c,\hat c^\dagger\right]=1$ and
$\left[\hat b,\hat c\right]=\left[\hat b^\dagger,\hat c^\dagger\right]=0$.
From the definitions given above, the Hamiltonian can be expressed in the compact form
of a quantum oscillator
\begin{equation}
\hat H=\hbar \omega\left(\hat c^\dag\hat c+\frac{1}{2}\right).\label{ham:ll}
\end{equation}
The number operators $\hat b^\dag\hat b$ and $\hat c^\dag\hat c$
commute and therefore
have simultaneous eigenstates. We therefore use their eigenvalues to label the kets
$\left\vert l,n\right\rangle$, where
$\hat c^\dag\hat c\left\vert l,n\right\rangle =l\left\vert l,n\right\rangle$ and
$\hat b^\dag\hat b\left\vert l,n\right\rangle =n\left\vert l,n\right\rangle$.
Additionally, from the properties of quantum oscillators,
we know that  $l,n=0,1,2\dots \infty$ .
Given that $\hat b$ and $\hat c$ follow the standard commutation rules of
quantum oscillators, we can take advantage of their properties as
rising and lowering operators and
express any normalized state of the charged particle as
\begin{equation}
\left\vert l,n \right\rangle
=\frac{\left(c^\dag\right)^l\left(\hat b^\dag\right)^n}{\sqrt{l! \,n!}}\left\vert 0,0 \right\rangle,
\label{ket:ln}
\end{equation}
where $\left\vert 0,0 \right\rangle$ is the ground-state. These are the Fock-Darwin states.
They are degenerate energy eigenstates with energy eigenvalues given by $E_l=\hbar\omega(l+1/2)$.
Here it is clear that $l$ tags the Landau levels and $n$ parametrizes the degeneracy.
The angular momentum may be conveniently expressed in terms of the rising and
lowering operators (\ref{op:b})-(\ref{op:ct}) as
\begin{equation}
\hat l_z=
\hat c^\dag\hat c-\hat b^\dag\hat b.
\end{equation}
Since $\hat l_z$ depends on the number operators,
the state $\left\vert l,n \right\rangle$ is also an eigenstate of the angular momentum,
i.e. $\hat l_z\left\vert l,n \right\rangle
=m\left\vert l,n \right\rangle=(l-n)\left\vert l,n \right\rangle$ where $m$ is the
angular momentum eigenvalue.
As $n$ and $l$ are positive integers, the angular momentum $m$ might be negative,
in contrast to the classical case, where the angular momentum is
always positive.
One of the most striking implications of this fact is
the existence of diamagnetic states  with negative OAM.

The very well known position-space wave function of Fock-Darwin states
is given by
\begin{multline}
\psi_{l,n}^{LG}\left(z,z^*\right)
=\frac{1}{\sqrt{2 \pi}l_B}\exp\left(-\frac{zz^*}{4l_B^2}\right)
\\\times\begin{cases}
(-1)^{l}\sqrt{\frac{l!}{n!}}\left(\frac{z^*}{\sqrt{2}l_B}\right)^{n-l}
L_l^{n-l}\left(\frac{zz^*}{2l_B^2}\right), & n\geq l,\\
(-1)^{n}\sqrt{\frac{n!}{l!}}\left(\frac{z}{\sqrt{2}l_B}\right)^{l-n}
L_n^{l-n}\left(\frac{zz^*}{2l_B^2}\right), & l\geq n,\\
\end{cases}\label{fockdarwin:psi}
\end{multline}
where  $z=x+iy$.
The corresponding probability density is given by
\begin{multline}
\rho^{LG}_{l,n}\left(r\right)
=\frac{1}{2 \pi l_B^2}\frac{\mathrm{min}(l,n)!}{\mathrm{max}(l,n)!}
\left(\frac{r^2}{2l_B^2}\right)^
{\left\vert l-n \right\vert}
\\
\times
\exp\left(-\frac{r^2}{2l_B^2}\right)
\left[L_{\mathrm{min}(l,n)}^{\left\vert l-n\right\vert}
\left(\frac{r^2}{2l_B^2}\right)\right]^2,
\end{multline}
where $z=r \exp(i \theta)$ and  $r^2=x^2+y^2=\left\vert z\right\vert^2$ .
These are precisely the LG modes.

\begin{figure*}
\includegraphics[angle=0,width=0.9 \textwidth]{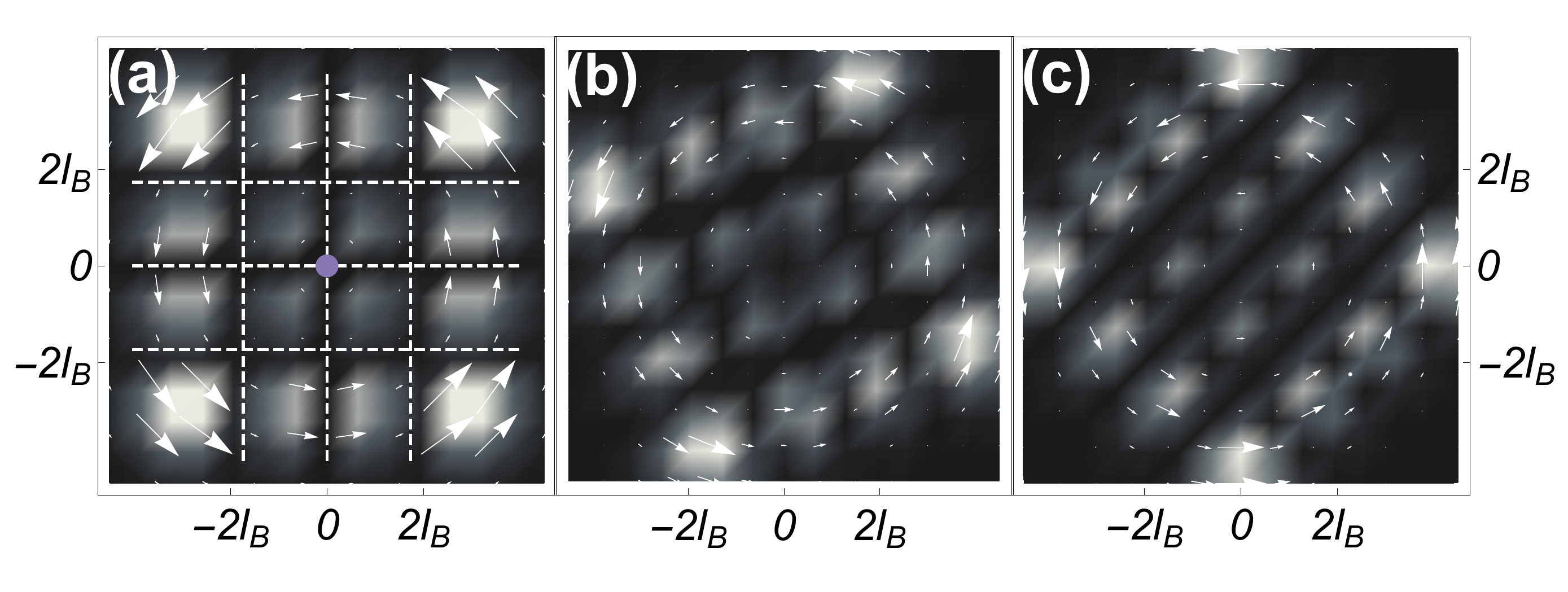}
\caption{Rotational dynamics of $\psi^{HG}_{3,3}$. A vector field
of the kinetic density current is superimposed to
the density plot to the probability density $\rho^{HG}_{3,3}$.
In (a) we observe the probability density $\rho^{HG}_{3,3}$ at $t=0$.
The dashed lines and the (purple) point indicate the position of the
singularities $S^{HG}_{3,3}$. The Figs. (b) and (c) show the evolution
of the probability density at $t=\pi/4\omega$ and
$t=\pi/2\omega$ respectively.}
\label{figure1}
\end{figure*}

\begin{figure}
\includegraphics[angle=0,width=0.5 \textwidth]{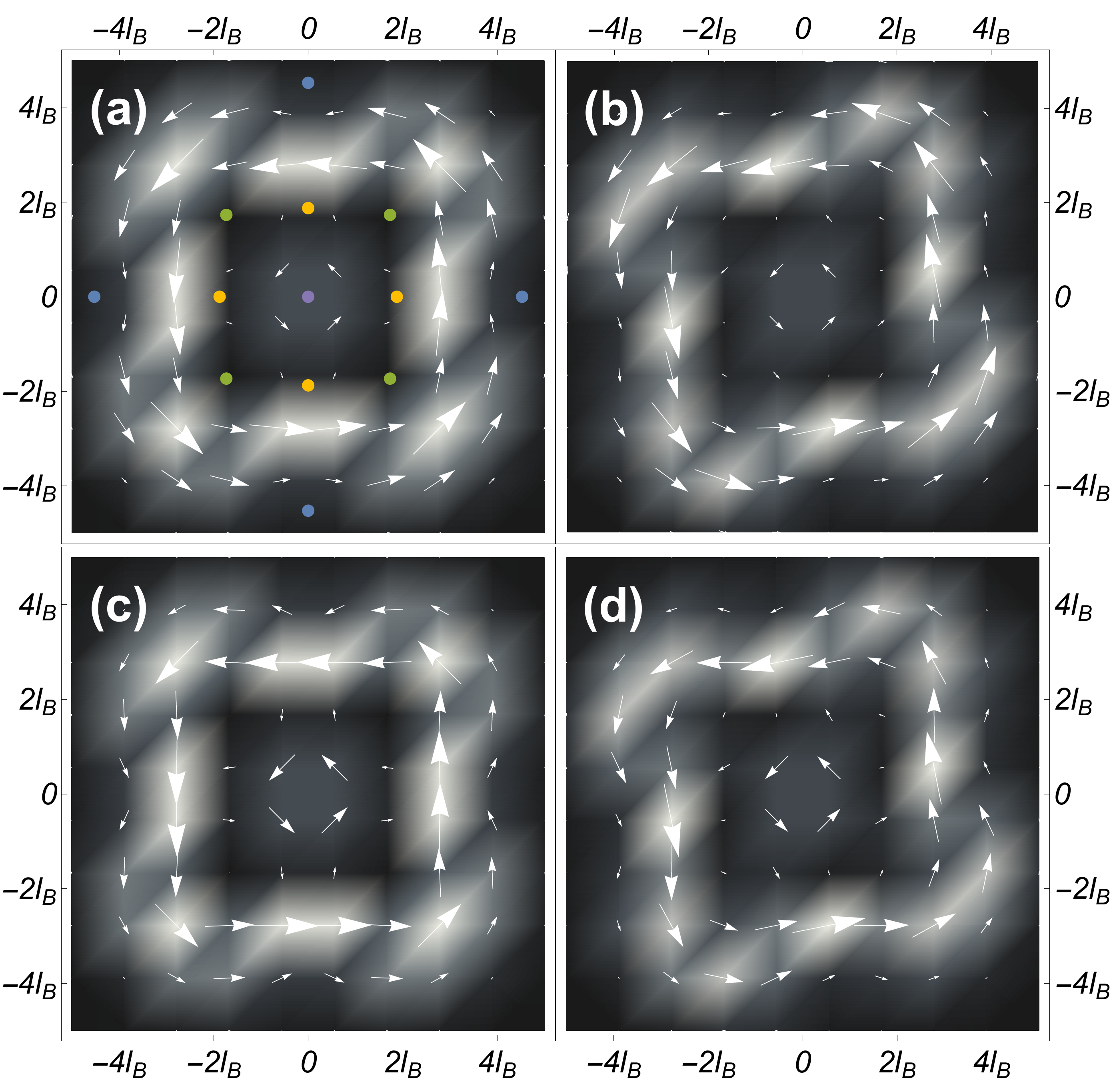}
\caption{Rotational dynamics of $\psi^{+HG}_{3,3}$.
Along the first row, in (a) and (b) we observe the time evolution of
$\rho^{+HG}_{3,3}$ and the vector field $\boldsymbol{J}^K$
for $t=0$ and $t=\pi/8\omega$ respectively.
In the second row, (c) and (d) show
the time evolution of
$\rho^{+HG}_{3,3}$ and the vector field $\boldsymbol{J}^C$
for $t=0$ and $t=\pi/8\omega$ respectively.
The positions of the different kinds of singularities are
indicated with (colored) dots in (a).}
\label{figure2}
\end{figure}

\begin{figure}
\includegraphics[angle=0,width=0.5 \textwidth]{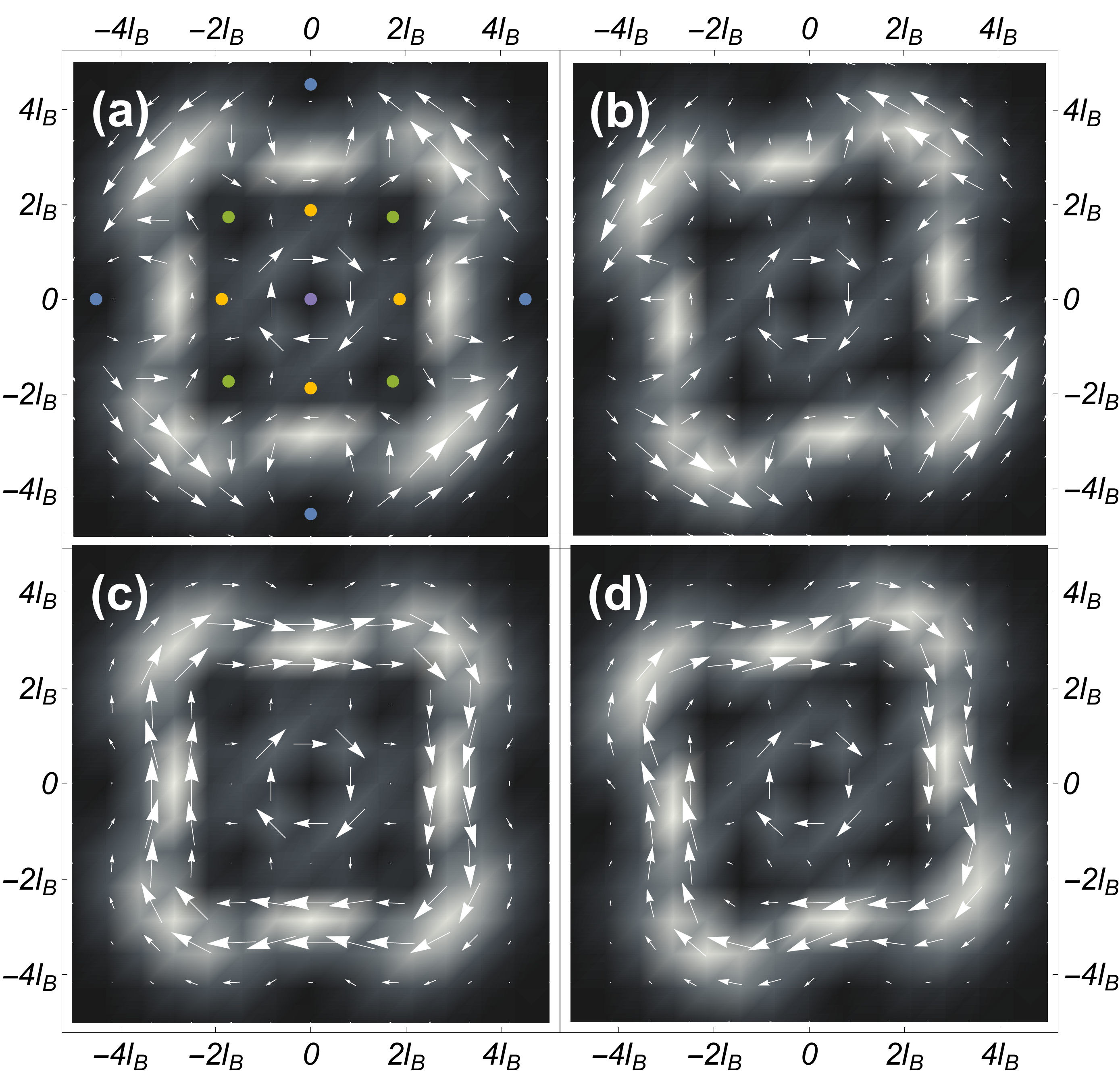}
\caption{Internal rotational dynamics of $\psi^{-HG}_{3,3}$.
Along the first row, in (a) and (b) we observe the time evolution of
$\rho^{-HG}_{3,3}$ and the vector field $\boldsymbol{J}^K$
for $t=0$ and $t=\pi/8\omega$ respectively.
In the second row, (c) and (d) show
the time evolution of
$\rho^{-HG}_{3,3}$ and the vector field $\boldsymbol{J}^C$
for $t=0$ and $t=\pi/8\omega$ respectively.
The positions of the different kinds of singularities are
indicated with (colored) dots in (a).}
\label{figure3}
\end{figure}

\begin{figure}
\includegraphics[angle=0,width=0.5 \textwidth]{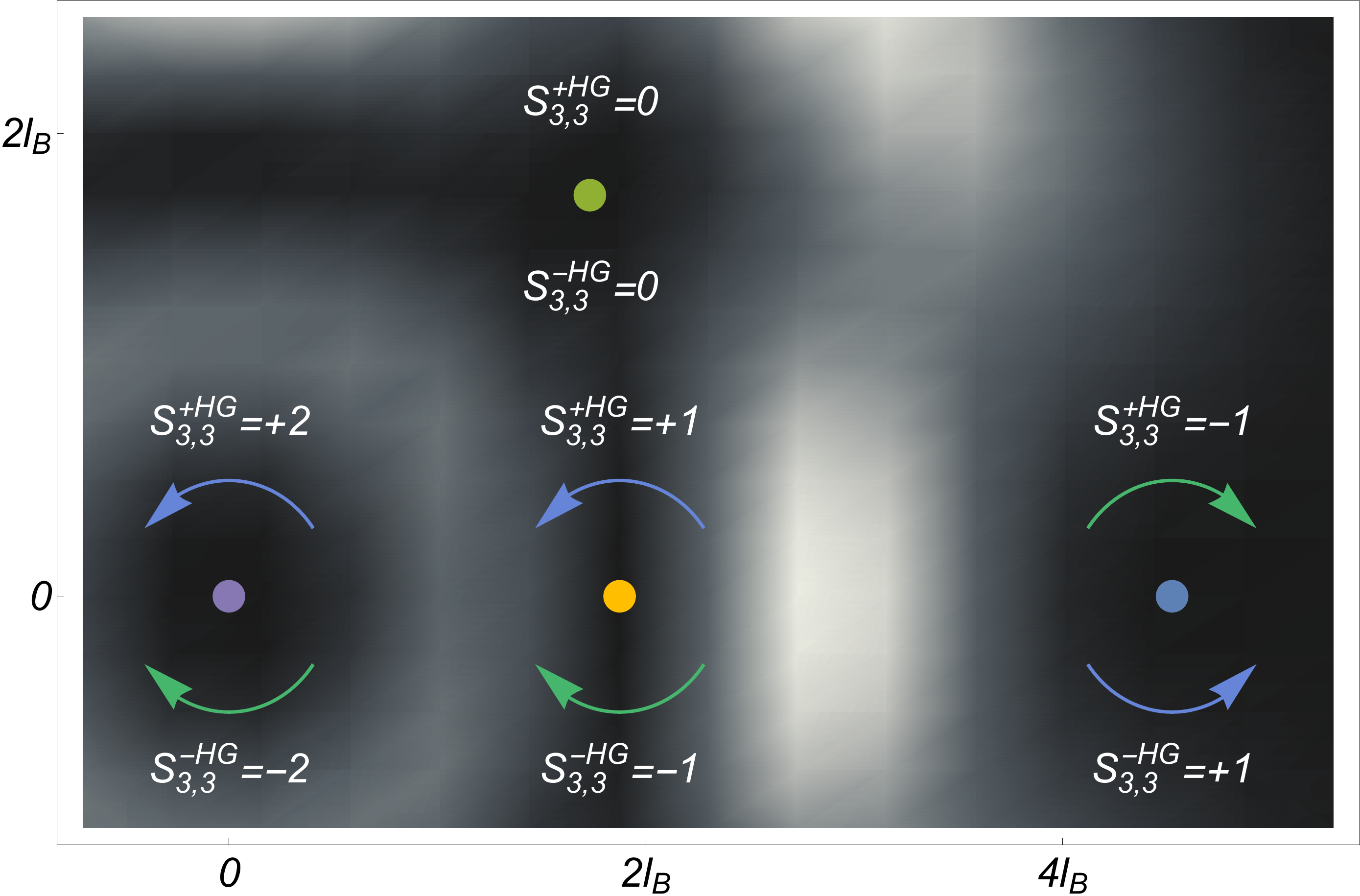}
\caption{Position and topological charges of the
$\psi^{+HG}_{3,3}$ and $\psi^{-HG}_{3,3}$ singularities.
The position and spinning directions of the main types of singularities
are superimposed to
the density plot to the probability density $\rho^{\pm HG}_{3,3}$.
These singularities are also shown in
Figs. \ref{figure2} and \ref{figure1}.}
\label{figure4}
\end{figure}

\begin{figure}
\includegraphics[angle=0,width=0.5 \textwidth]{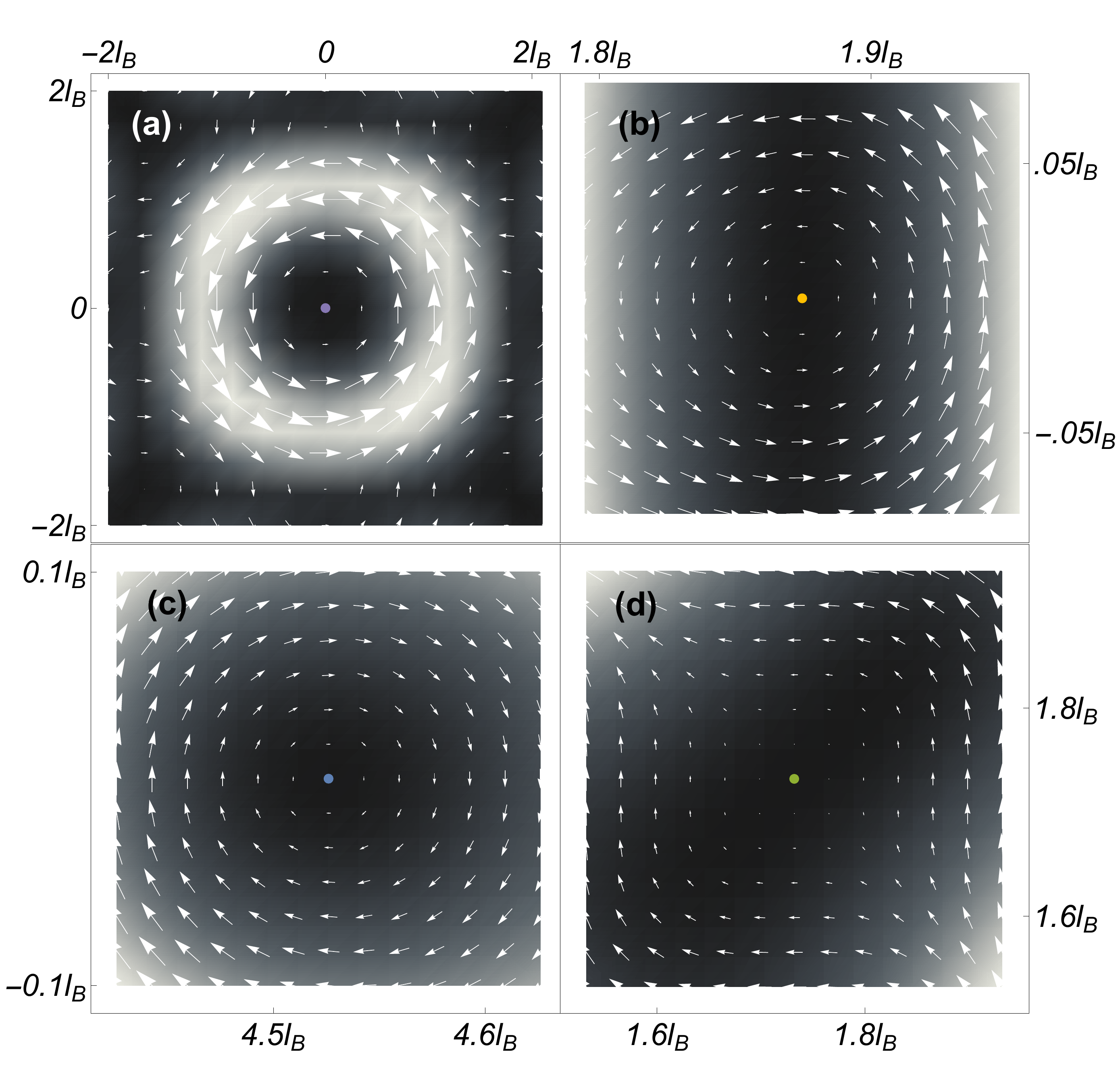}
\caption{ Singularities of $\psi^{+HG}_{3,3}$. The positions
of the four main types
of $\psi^{+HG}_{3,3}$ singularities are indicated with (colored) dots.
The density plot of the density probability $\rho^{+HG}_{3,3}$
along with the corresponding vector field $\boldsymbol{J}^C$ are shown.
(a), (b), (c) and (d) show the vorticity
of the $S^{+HG}_{3,3}=+2$, $S^{+HG}_{3,3}=+1$,
$S^{+HG}_{3,3}=-1$ and $S^{+HG}_{3,3}=0$
topological charges, respectively.
 }
\label{figure5}
\end{figure}

\begin{figure}
\includegraphics[angle=0,width=0.5 \textwidth]{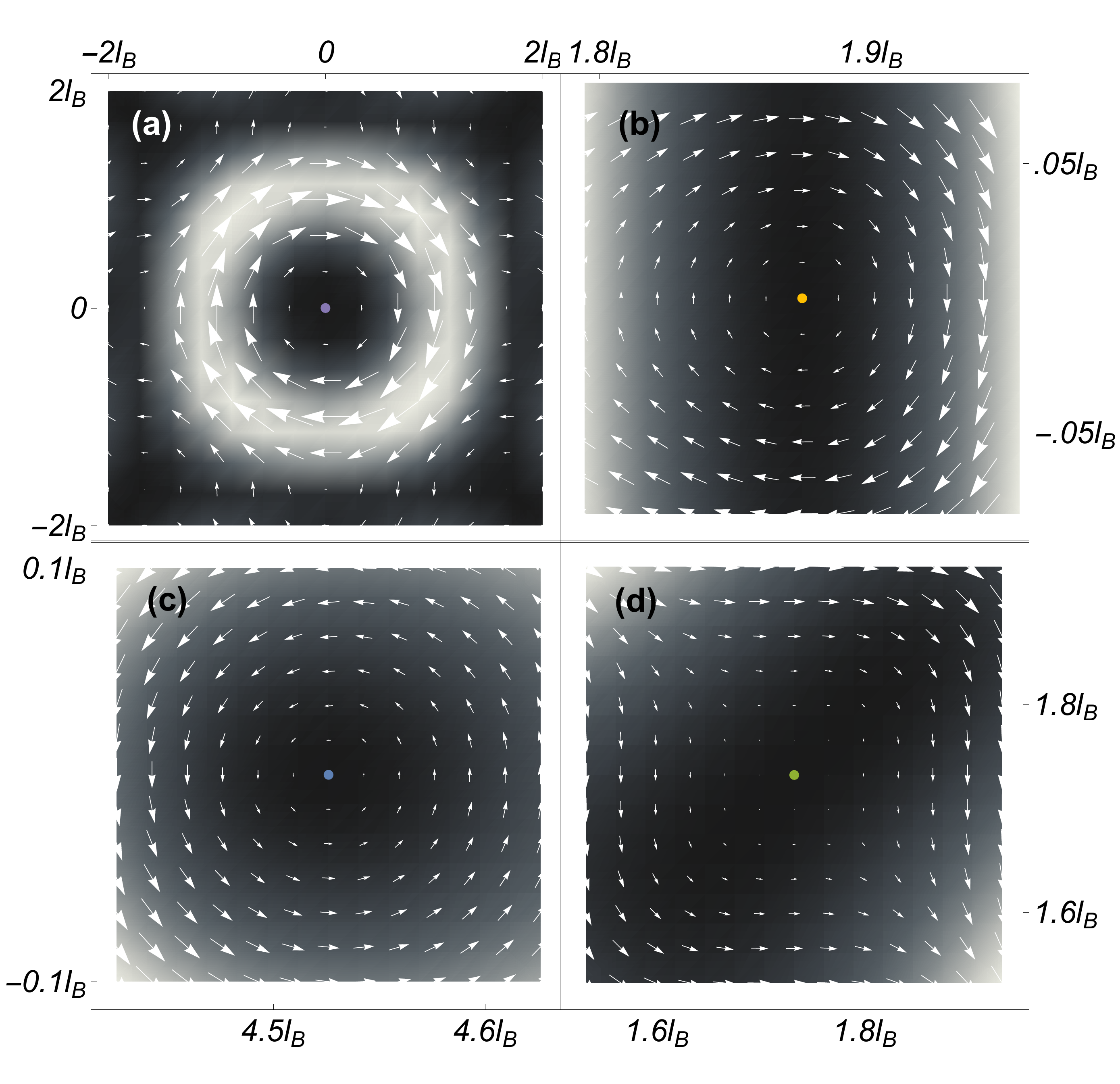}
\caption{ Singularities of $\psi^{-HG}_{3,3}$. The positions
of the four main types
of $\psi^{-HG}_{3,3}$ singularities are indicated with (colored) dots.
The density plot of the density probability $\rho^{-HG}_{3,3}$
along with the corresponding vector field $\boldsymbol{J}^C$ are also shown.
(a), (b), (c) and (d) show the vorticity
of the $S^{-HG}_{3,3}=-2$, $S^{-HG}_{3,3}=-1$,
$S^{-HG}_{3,3}=+1$ and $S^{-HG}_{3,3}=0$
topological charges, respectively.}
\label{figure6}
\end{figure}

\begin{figure}
\includegraphics[angle=0,width=0.5 \textwidth]{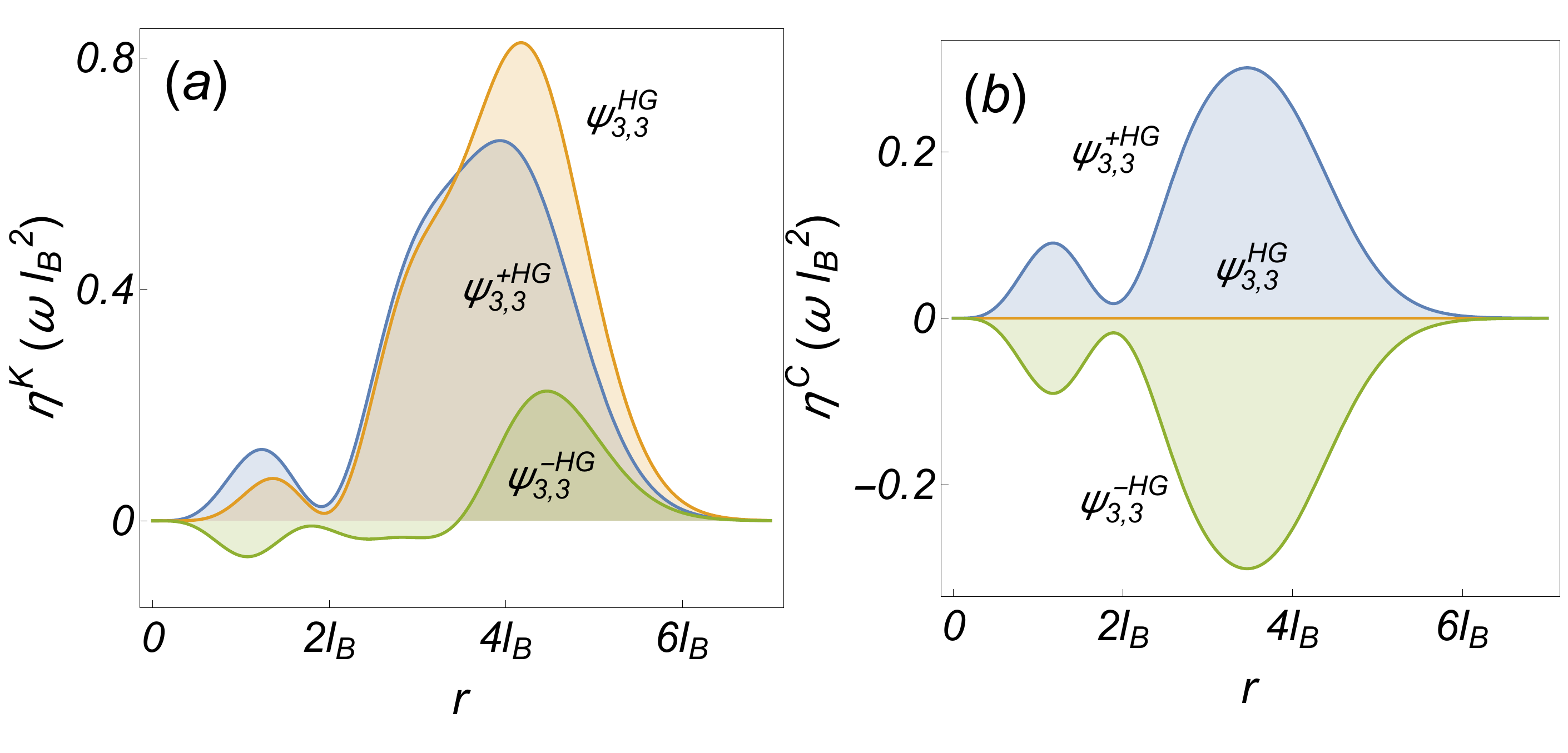}
\caption{ $\eta^K$ and $\eta^C$ for $\psi^{HG}_{3,3}$ (orange),
$\psi^{+HG}_{3,3}$ (blue) and $\psi^{-HG}_{3,3}$ (green) as functions
of $r=\sqrt{x^2+y^2}$.
In the negative beam $\psi^{-HG}_{3,3}$ $\eta^K$ 
changes sign and $\eta^C$
is symmetric with respect to the positive beam $\psi^{+HG}_{3,3}$. }
\label{figure7}
\end{figure}

\begin{figure}
\includegraphics[angle=0,width=0.5 \textwidth]{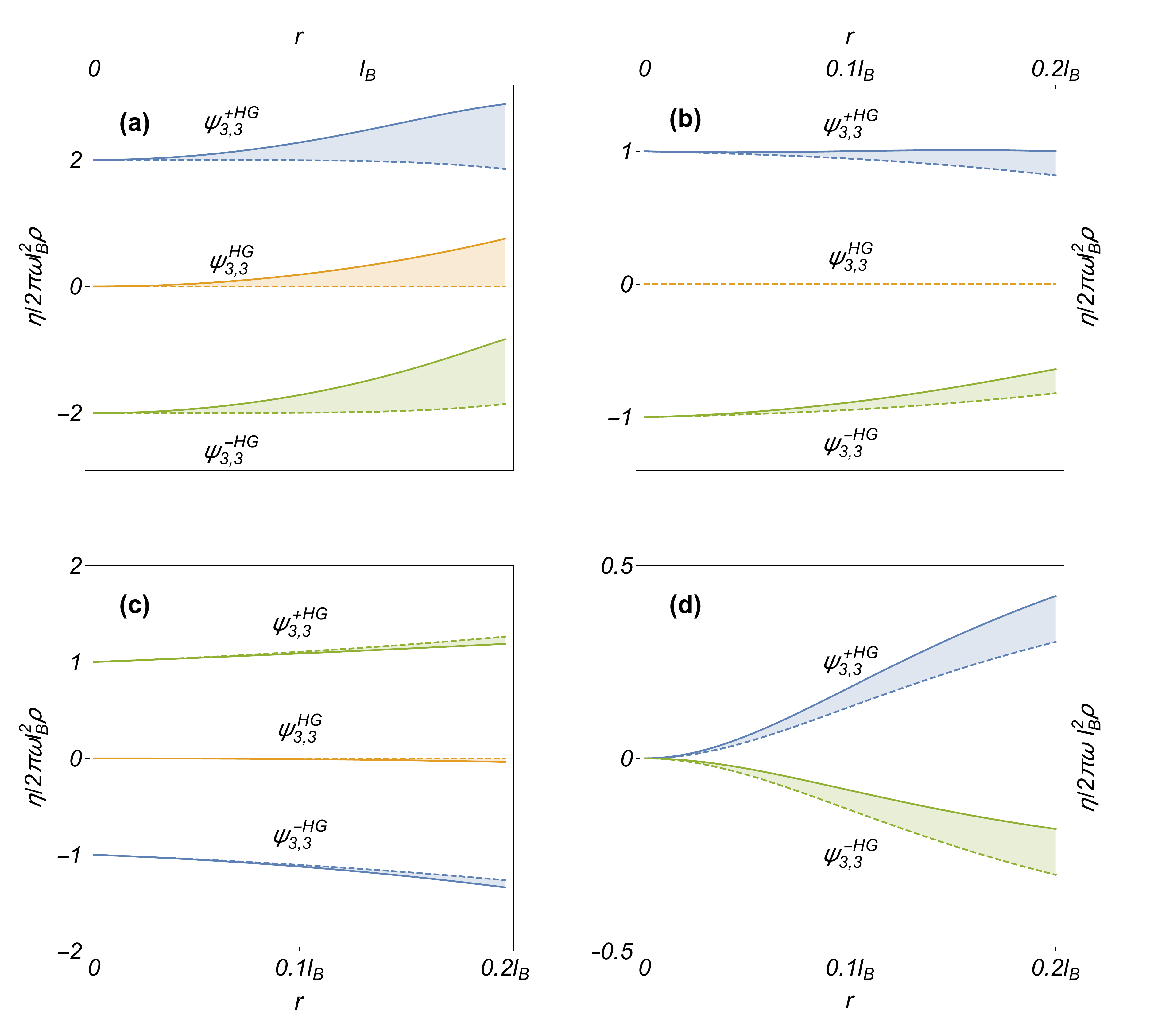}
\caption{ Topological charges of the $\psi^{-HG}_{3,3}$
and $\psi^{+HG}_{3,3}$ singularities. 
$\eta^K/2\pi \omega l_B^2\rho$ (solid lines) and
$\eta^C/2\pi \omega l_B^2\rho$ (dashed lines)
are ploted as functions of $r$, the distance to the singularity.
(a), (b), (c) and (d) show $\eta^C/2\pi l_B^2\rho$ and
$\eta^K/2\pi l_B^2\rho$
corresponding to the $S^{\pm HG}_{3,3}=\pm 2$, $S^{\pm HG}_{3,3}=\pm 1$,
$S^{\mp HG}_{3,3}=\mp 1$ and $S^{\pm HG}_{3,3}=0$
singularities respectively.
The singularities (a), (c) and (d) are approached along the
angle $\theta=\pi/3$,
and the one in (b) is approached along $\theta=0$.}
\label{figure8}
\end{figure}

\section{Superposition of LG modes.}\label{compbeams}
Any beam can in principle be expressed as a linear superposition
of LG mode as
\begin{equation}
\psi(x,y)=\sum_{l,n}A_{l,n}\psi^{LG}_{l,n}(x,y),\label{psi:gen}
\end{equation}
where
\begin{equation}
A_{l,n}=\left\langle l,n\big| \psi\right\rangle=\int dxdy \psi^{LG*}_{l,n}(x,y) \psi(x,y).
\end{equation}
A particular case are HG beams.
HG electron beams have been generated by means of
mode converters \cite{PhysRevLett.109.084801} that add or subtract units of
the topological charge to an incident beam.
Therefore a LG mode having non-vanishing angular momentum and topological
charge can be turned into a HG mode with vanishing topological charge. 
An immediate application of mode converters would then be mode discrimination.
Adding one charge unit to a given negatively charged
vortex would allow to distinguish it from
its positively charged twin by comparing the otherwise
identical spatial intensity distributions \cite{PhysRevLett.109.084801}.

HG beams can be thought of as 
stationary solutions of
a two-dimensional isotropic oscillator. Their wave function is given by
\begin{multline}
\psi^{HG}_{j,k}\left(x,y\right)=\frac{1}{\sqrt{2\pi 2^j2^kj!k!}l_B}
\exp\left(-\frac{x^2+y^2}{4l_B^2}\right)\\
\times H_j\left(\frac{x}{\sqrt{2}l_B}\right)
H_k\left(\frac{y}{\sqrt{2}l_B}\right).
\end{multline}

Using the 
relation between Hermite and Laguerre 2D
polynomials\cite{Wunsche2001665}
\begin{multline}
H_j(x)H_k(y)=i^k\sum_{q=0}^{j+k}2^qP_q^{\left(j-q,k-q\right)}\left(0\right)\\
\times
\begin{cases}
(-1)^{q}q!(x-iy)^{j+k-2q} & \\
\,\,\,\,\,\,\,\,\,\,\,\,\,\,\,\,\,\,
\,\,\,\,\,\,\,\,\,\,\,\,\,\,\,\,\,\,
\times L^{j+k-2q}_{q}\left(x^2+y^2\right), & 2q \leq j+k, \\
(-1)^{j+k-q} (j+k-q)!(x+iy)^{2q-j-k} & \\
\,\,\,\,\,\,\,\,\,\,\,\,\,\,\,\,\,\,
\,\,\,\,\,\,\,\,\,\,\,\,\,\,\,\,\,\,
\times L^{2q-j-k}_{j+k-q}\left(x^2+y^2\right), &
2q \geq j+k,\\
\end{cases}
\end{multline}
one can identify the $A_{l,n}$ coefficients
for HG beams.
Their wave functions adopt the form
\begin{equation}
\psi^{HG}_{j,k}\left(x,y\right)=\sum_{q=0}^{j+k}A^{j,k}_q
\psi^{LG}_{q,j+k-q}\left(x,y\right),\label{HG:stationary}
\end{equation}
where
\begin{equation}
A^{j,k}_q=2^qi^k\sqrt{\frac{q!\left(j+k-q\right)!}{2^j2^kj!k!}}
P_q^{\left(j-q,k-q\right)}\left(0\right),\label{HG:stationarycoef}
\end{equation}
and $P_q^{j-q,k-q}\left(0\right)$ are the Jacobi polynomials.
From Eq. (\ref{HG:stationary}) we can readily calculate the
time-dependent HG states
\begin{multline}
\psi^{HG}_{j,k}\left(x,y,t\right)\\
=\sum_{q=0}^{j+k}A^{j,k}_q
\exp\left[-i\omega \left(q+\frac{1}{2}\right) t\right]
\psi^{LG}_{q,j+k-q}\left(x,y\right).\label{HG:evolve}
\end{multline}

In general, HG beams have the structure of balanced
states \cite{PhysRevX.2.041011} i.e. they are formed of linear combinations of 
LG beams where states having opposite angular momentum
participate with the same weight. Thereby  HG beams have
vanishing overall angular momentum.
However, it is possible to separate HG beams
in a negative, vanishing and positive part as
\begin{equation}
\psi_{i,j}^{HG}=\psi_{i,j}^{-HG}+\psi_{i,j}^{0HG}+\psi_{i,j}^{+HG}.
\end{equation}
These are correspondingly formed of LG states having negative,
vanishing and positive angular momenta.
Nevertheless, they are not necessarily angular momentum
eigenstates.
In pure HG beams phase singularities are arranged as lines with vanishing
topological charge.
Conversely, the negative and positive parts of the HG
states possess phase singularities arranged as isolated points with different
topological charges.
We are particularly interested in these states because of  their rich and
complex singularity structure.

The vanishing part of an  HG beam is
given by
\begin{multline}
\psi^{0HG}_{j,k}\left(x,y,t\right)
=A^{j,k}_{(j+k)/2}\exp\left(-i \frac{j+k}{2}\omega t\right)
\\
\times
\begin{cases}
\psi^{LG}_{(j+k)/2,(j+k)/2}\left(x,y\right), & i \,\mathrm{and} \, j \, \mathrm{even \,numbers},\\
0, & i \, \mathrm{or} \, j \, \mathrm{odd \, numbers}.\\
\end{cases}
\end{multline}
The positive and negative parts can be expressed as
\begin{equation}
\psi^{\pm HG}_{j,k}\left(x,y,t\right)
=\sum_{q=q^{\pm}_{min}}^{q^{\pm}_{max}}A^{j,k}_q
\mathrm{e}^{-iq\omega t}
%\exp(-iq\omega t)
\psi^{LG}_{q,j+k-q}\left(x,y\right),
\end{equation}
where
\begin{eqnarray}
q^-_{min} &=& 0,\label{q:m:min}\\
q^-_{max} &=& \mathrm{int}\left( \frac{j+k}{2}\right)-p,\\
q^+_{min} &=& \mathrm{int}\left( \frac{j+k}{2}\right)+p,\\
q^+_{max} &=& j+k,\label{q:p:max}
\end{eqnarray}
with $p=1$ if $i$ and $j$ are even, and $p=0$ if $i$ or $j$ are odd.
Pure HG beams are generated
for $q_{min}=0$ and $q_{max}=j+k$.

Other superpositions of LG modes, as the ones given by
\begin{eqnarray}
\psi_n^{+BEC} &=& \frac{1}{\sqrt{5}}\psi^{LG}_{0,0}+\frac{2}{\sqrt{5}}\psi^{LG}_{n,0},
\label{plusbec}\\
\psi_n^{-BEC} &=& \frac{1}{\sqrt{5}}\psi^{LG}_{0,0}+\frac{2}{\sqrt{5}}\psi^{LG}_{0,n},
\label{minusbec}
\end{eqnarray}
yield interesting singularity arrangements.
It has been proved that their optical analogues have $n$
vortices  located at equally spaced points around the center.
Their vortex structure has been studied
in optical beams \cite{PhysRevLett.88.013601}
and plays an important role in
Bose-Einstein condensates stirred with
a laser beam \cite{PhysRevLett.84.806,PhysRevX.2.041011}.

\begin{figure}
\includegraphics[angle=0,width=0.5 \textwidth]{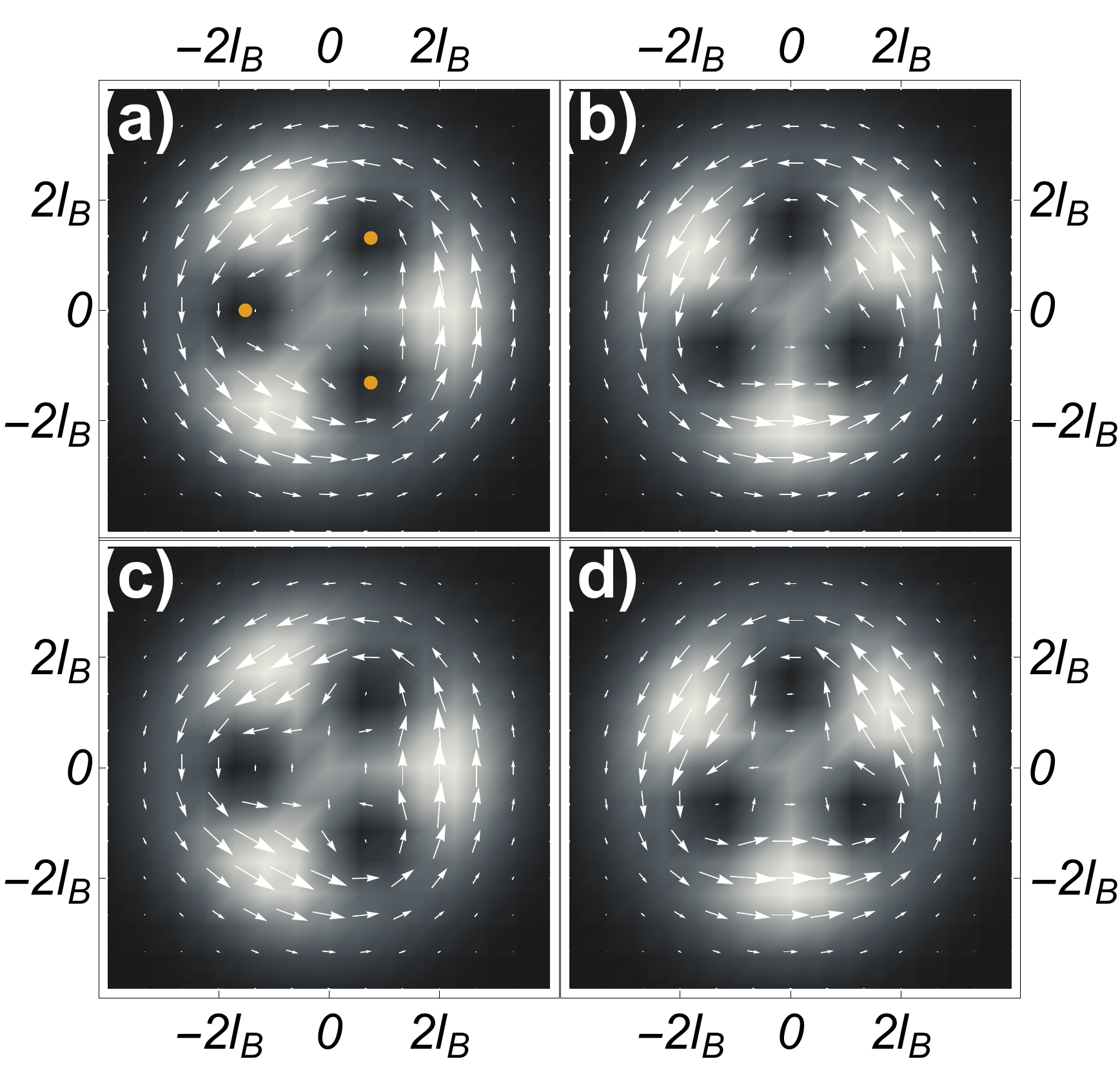}
\caption{Rotational dynamics of $\psi^{+BEC}_{3}$.
Along the first row, in (a) and (b) we observe the time evolution of
$\rho^{+BEC}_{3}$ and the vector field $\boldsymbol{J}^K$
for $t=0$ and $t=\pi/6\omega$ respectively.
In the second row, (c) and (d)  show
the time evolution of
$\rho^{+BEC}_{3}$ and the vector field $\boldsymbol{J}^C$
for $t=0$ and $t=\pi/6\omega$ respectively.
The positions of the three singularities are
indicated with (orange) dots in (a).}
\label{figure9}
\end{figure}

\begin{figure}
\includegraphics[angle=0,width=0.5 \textwidth]{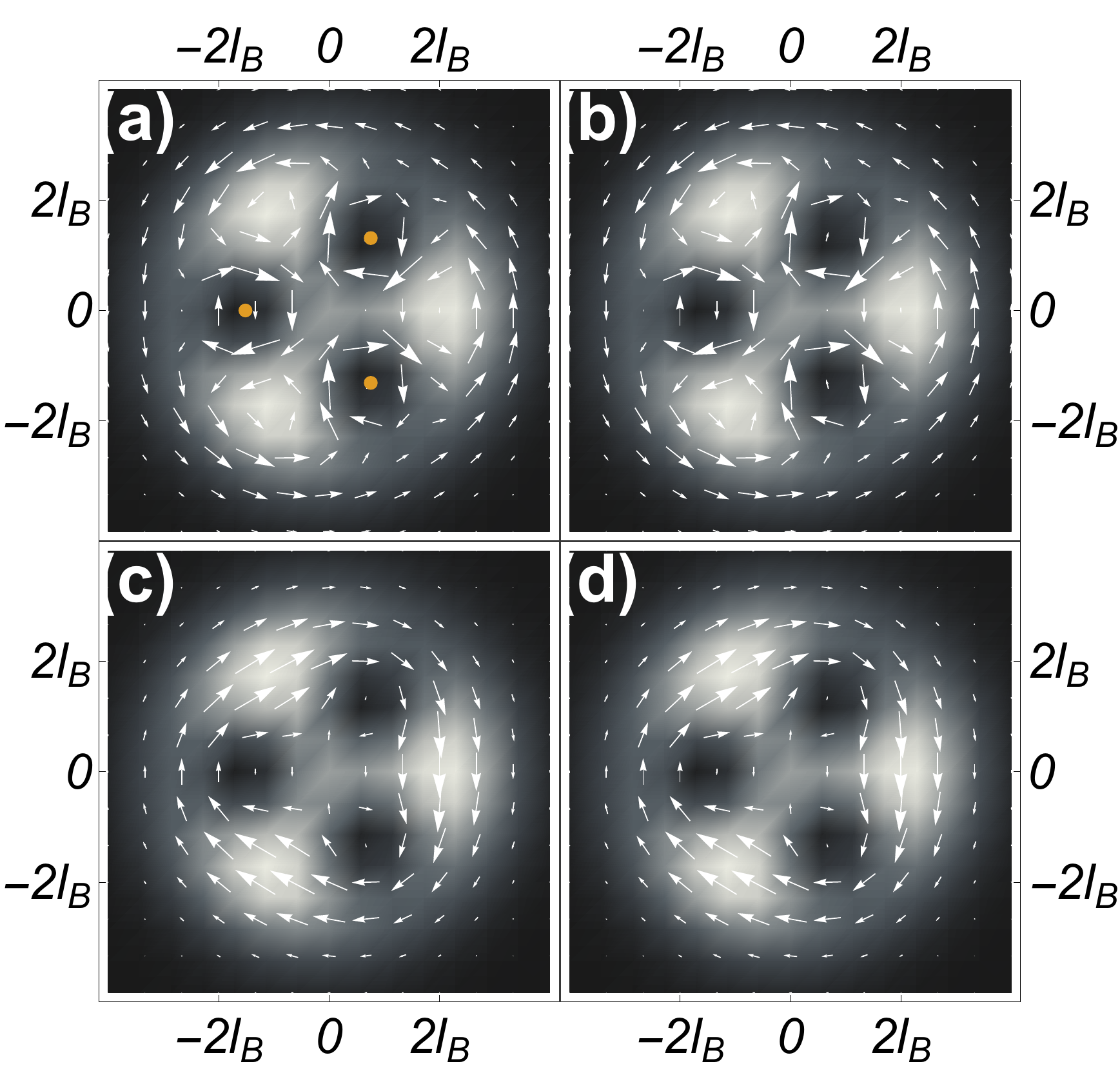}
\caption{ Rotational dynamics of $\psi^{-BEC}_{3}$.
Along the first row, in (a) and (b) we observe the time evolution of
$\rho^{-BEC}_{3}$ and the vector field $\boldsymbol{J}^K$
for $t=0$ and $t=\pi/6\omega$ respectively.
The singularities clearly revolve around the center in the counterclockwise direction.
In the second row, (c) and (d)  show
the time evolution of
$\rho^{-BEC}_{3}$ and the vector field $\boldsymbol{J}^C$
for $t=0$ and $t=\pi/6\omega$ respectively.
The positions of the three singularities are
indicated with (orange) dots in (a).}
\label{figure10}
\end{figure}

\begin{figure}
\includegraphics[angle=0,width=0.5 \textwidth]{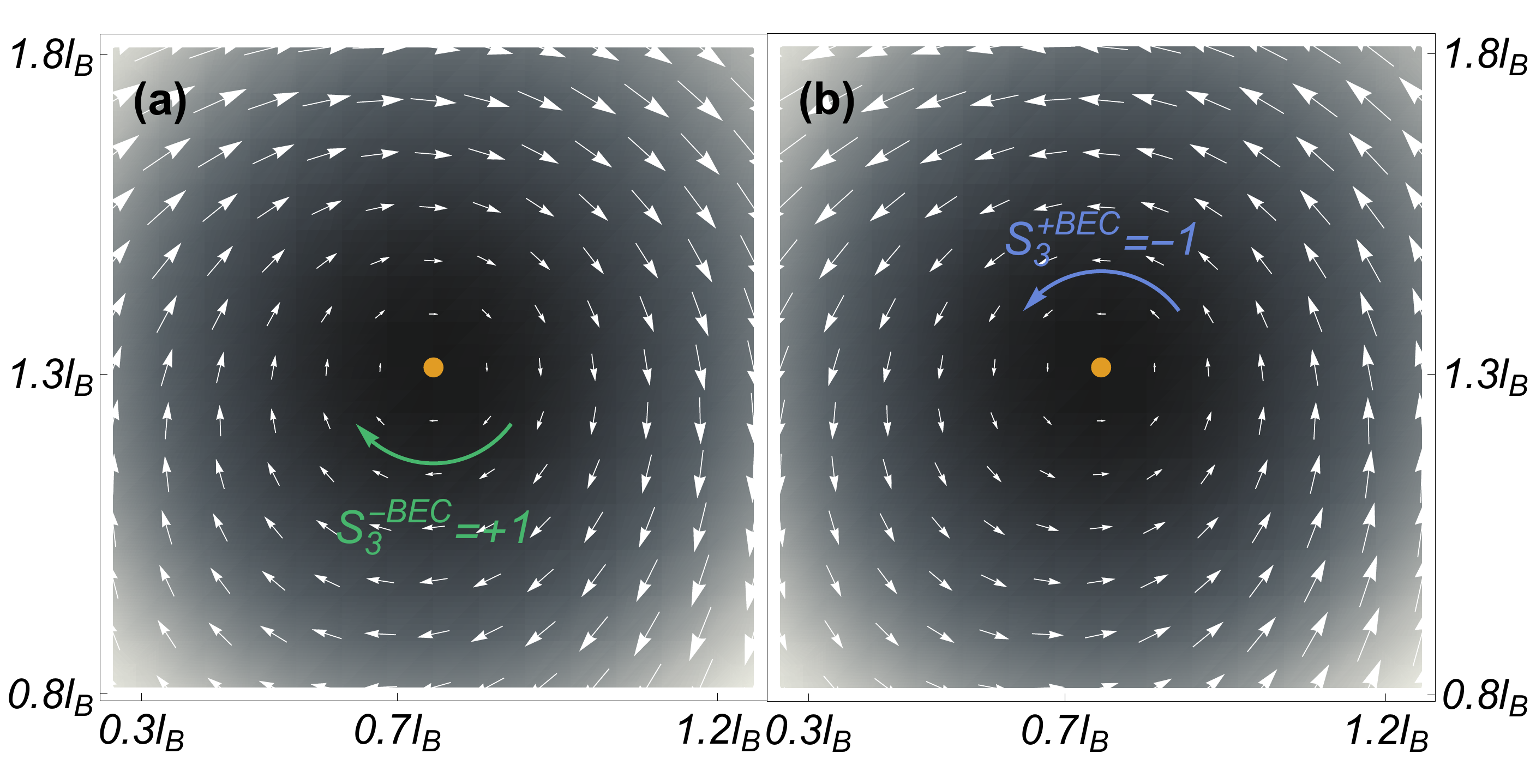}
\caption{ Topological charges and positions of the
(a) $\psi^{+BEC}_{3}$ and (b) $\psi^{-BEC}_{3}$ singularities.
The position and the kinetic density currents' spinning direction of the singularities
are superimposed to
the density plot to the probability density $\rho^{\pm BEC}_{3}$.
The vector plot indicates the vector field of the kinetic density current $\boldsymbol{J}^K$.
These singularities are also shown in
Figs. \ref{figure9} and \ref{figure10}.}
\label{figure11}
\end{figure}

\begin{figure}
\includegraphics[angle=0,width=0.5 \textwidth]{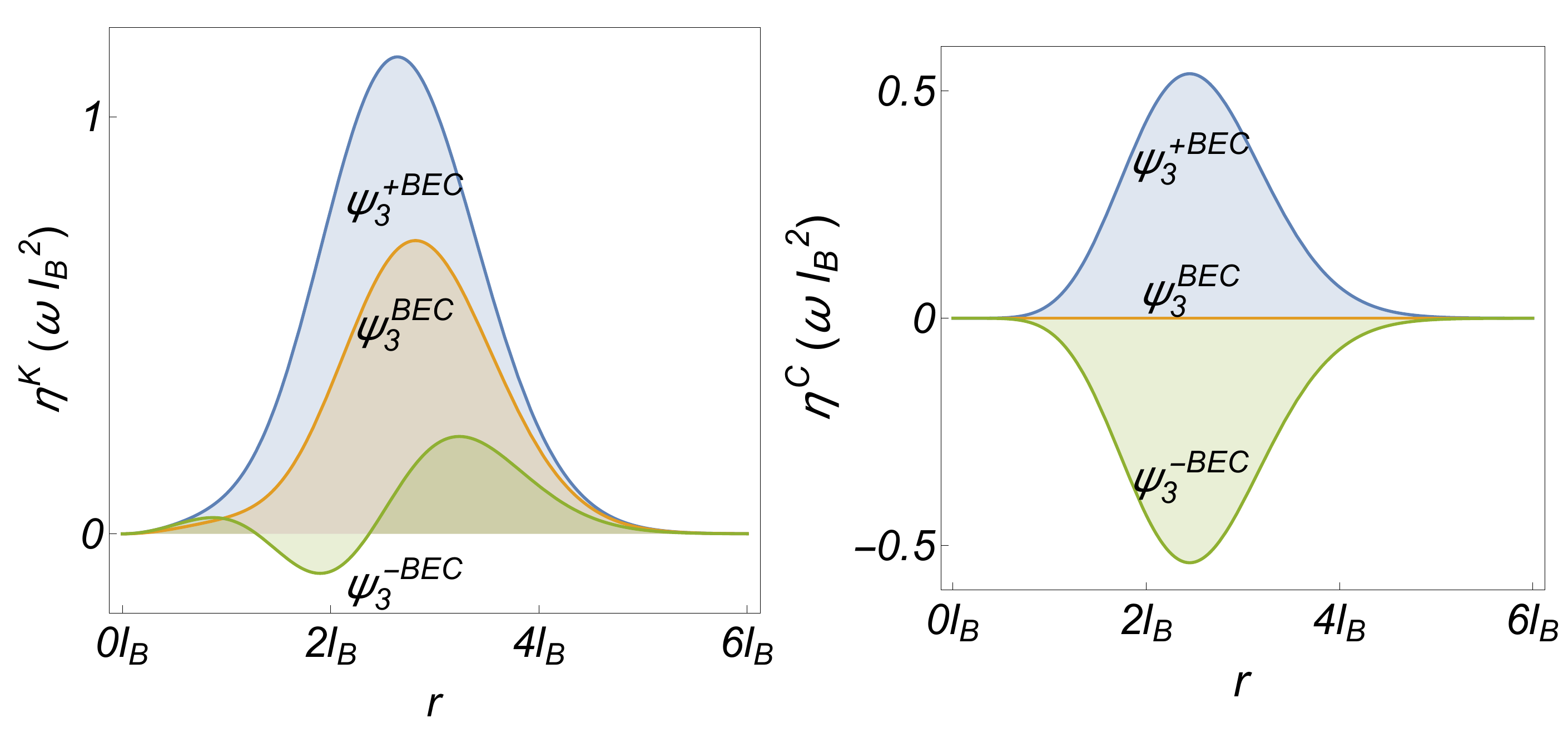}
\caption{ $\eta^K$ and $\eta^C$ for $\psi^{BEC}_{3}$ (orange),
$\psi^{+BEC}_{3}$ (blue) and $\psi^{-BEC}_{3}$ (green) as functions
of $r=\sqrt{x^2+y^2}$.
In the negative beam $\psi^{-BEC}_{3}$ $\eta^K$ 
changes sign and $\eta^C$
is symmetric with respect to the positive beam $\psi^{+BEC}_{3}$.}
\label{figure12}
\end{figure}

\begin{figure}
\includegraphics[angle=0,width=0.5 \textwidth]{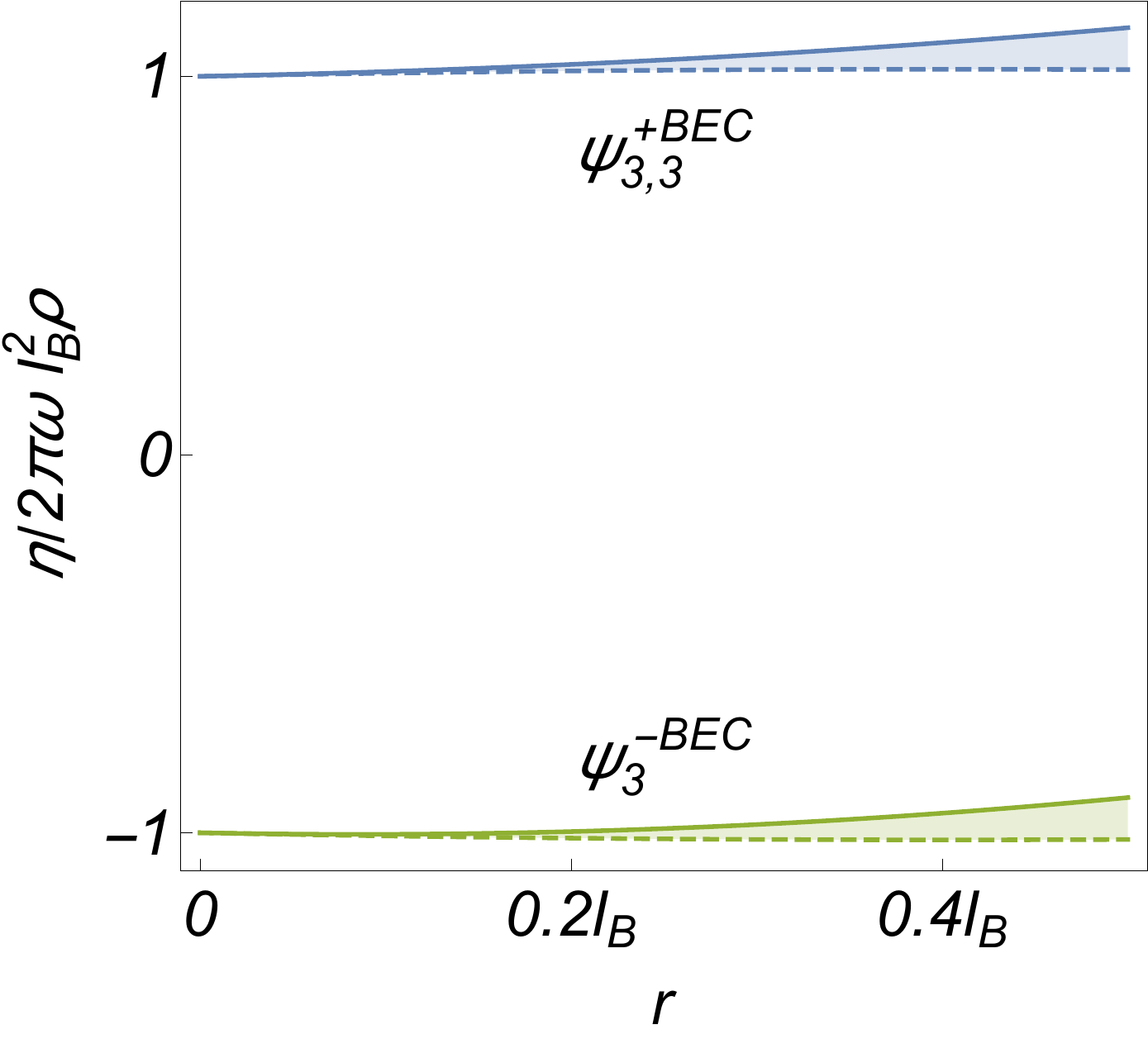}
\caption{ Topological charges of the $\psi^{-BEC}_{3}$
and $\psi^{+BEC}_{3}$ singularities. 
$\eta^K/2\pi \omega l_B^2\rho$ (solid lines) and
$\eta^C/2\pi \omega l_B^2\rho$ (dashed lines)
are plotted as functions of $r$, the distance to the singularity.
The plot is shown for
the $S^{\pm BEC}_{3}=\pm 1$ singularity located in $z=(2\sqrt{3})^{1/3}\exp(i \pi/3)$.
The singularity is approached along the
angle $\theta=0$.}
\label{figure13}
\end{figure}

\section{Topological charge of electron beams}\label{singu:struc}

Any point in the wave function's domain where the amplitude vanishes
presents a  singularity where the phase is undefined.
These are also called vortex singularities because
the wave function's phase circulates around these points.
The singularity of a state $\psi=\left\vert \psi \right\vert \mathrm{e}^{i\phi}$
may be characterized by its topological charge
\begin{equation}
S=\frac{1}{2\pi}\oint_{\mathcal{C}}\nabla \phi\cdot d\boldsymbol{l}
=\frac{1}{2\pi}\int_{0}^{2\pi}\frac{d\phi}{d \theta}d\theta,
\label{topcharge:gen}
\end{equation}
where $\mathcal{C}$ is a closed path around a point
where $\psi=\left\vert \psi\right\vert \exp(i\phi)$ vanishes.

In the particular case of LG beams
the calculation of the topological charge is straightforward.
These states only possess one singularity located in the
symmetry axis of the beam ($z=0$).
It can be easily verified that the Fock-Darwin
states in Eq. (\ref{fockdarwin:psi}) have an $(n-l)$-fold zero in $z=0$.
In this point the LG modes vanish if $l\ne n$ and
consequently the phase is undefined. 
By setting $z=r\exp(i\theta)$ in Eq. (\ref{fockdarwin:psi})
we notice that the wave function's phase is $\phi=(l-n)\theta$
and, from Eq. (\ref{topcharge:gen}), the topological charge is given by
\begin{equation}
S_{l,n}^{LG}=(l-n).
\end{equation}
Thus, the topological charge of the singularity in LG states
is identical to the angular momentum.
The structure of the LG modes
in the vicinity of the singularity
are listed in rows 1-2 of Table \ref{singu:table}.

In beams whose wave functions are not eigenstates
of the angular momentum, singularities may have
a more involved structure.
An arbitrary state $\psi$ may have a number of 
phase singularities
$z_1=x_1+iy_1$, $z_2=x_2+iy_2$, $\dots$, $z_n=x_n+iy_n$.
Expanding the wave function around one of them, say
$z_j$, up to the lowest non-trivial power we obtain
\begin{equation}
\psi=a\left[\gamma X^\prime
+i\beta Y^\prime \right]^s.
\label{singu:form}
\end{equation}
The coordinates $X=x-x_j$ and $Y=y-y_j$ belong to a frame oriented
along the singularity's symmetry axes  and whose origin
is located in the singularity. If the singularity's symmetry axes form
an $\nu$ angle with respect to the standard frame, then
$X^\prime = \cos \nu X-\sin\nu Y$, $Y^\prime = \cos \nu Y+\sin\nu X$
where $X^\prime$ and $Y^\prime$ are in the standard reference frame.
The $\gamma$ and $\beta$ parameters
are real and $a=\left\vert a\right\vert \exp(i\alpha)$ is
a complex number. The wave function 
will have elliptical symmetry and in general it will not be aligned with
the standard reference frame.
Although the wave function may have other forms
in the vicinity of a singularity, (\ref{singu:form}) is general enough
to allow the analysis of the wave functions presented here.

In order to perform a closed integral (\ref{topcharge:gen}) around the singularity and
since in general $\gamma \neq \beta$ it is convenient to set
\begin{equation}
\bar z-\bar z_j= \left\vert\gamma\right\vert (x-x_j)+i\left\vert\beta\right\vert (y-y_j)=r \exp(i\varphi).
\end{equation} 
Performing the closed integral around a circle in the $\bar{z}$ variable
($0\leq \varphi \leq 2\pi$)
is equivalent to integrating the original $z$ variables around an ellipse ($0\leq \theta \leq 2\pi$) 
centered in $z_j$ and whose
semi-major and semi-minor axes are $\gamma$ and $\beta$.

Three different cases must be considered: 1) $\gamma\beta >0$,
2) $\gamma\beta<0$ and 3) $\gamma\beta=0$.
If $\gamma\beta>0$ then the wave function (\ref{singu:form}) can be
approximated by
$\psi=\pm\left\vert a\right\vert \bar z^s=\pm\left\vert a\right\vert r^s\exp(i\alpha+s\varphi)$.
Therefore the wave function's phase in  $\bar{z}=\bar{z}_j$ is
$\phi=(1\mp1)\pi/2+\alpha+s\varphi$. Inserting the phase in the integral
(\ref{topcharge:gen}), the topological charge yields
\begin{equation}
S=\frac{1}{2\pi}\int d\phi= \frac{1}{2\pi}s\int_0^{2\pi} d\varphi=s, \,\,\,\,\,\,\,\,\, \gamma\beta>0.
\end{equation}
Similarly, if $\gamma\beta<0$ the singularity is located in the isolated point $\bar z^*=\bar z_j^*$.
Thereby the first non-vanishing term of the wave function expansion
is given by $\psi=\pm\left\vert a\right\vert (\bar z^*)^s=\pm\left\vert a\right\vert r^s\exp(i\alpha-s\varphi)$
and the topological charge yields
\begin{equation}
S=\frac{1}{2\pi}\int d\phi= -\frac{1}{2\pi}s\int_0^{2\pi} d\varphi=-s,
 \,\,\,\,\,\,\,\,\, \gamma\beta<0.
\end{equation}
We have quite a different situation when $\gamma\beta=0$.
In this case, the phase of the wave function close
to the singularity is constant. If $\gamma=0$ then
$\phi=\pi/2$ and if $\beta=0$ then $\phi=0$. In either case
the phase is constant and therefore
\begin{equation}
S=\frac{1}{2\pi}\int d\phi=0,
 \,\,\,\,\,\,\,\,\,
 \,\,\,\,\,\,\,\,\,
\gamma\beta=0.
\end{equation}

\section{Kinetic and canonical currents}\label{currentsKC}

In most quantum mechanical systems the canonical momentum $\hat{\boldsymbol{p}}$
is proportional to the velocity operator $\hat{\boldsymbol{v}}$.
However, when the effect of the magnetic field is introduced,
this is no longer true due to extra terms arising from the vector potential
$\boldsymbol{A}$. These terms strongly depend on the gauge choice.
One can then distinguish two different kinds of momenta:
the canonical momentum $\hat{\boldsymbol{p}}$ and the
kinetic momentum $m\hat{\boldsymbol{v}}$
\cite{1367-2630-17-9-093015,PhysRevLett.113.240404}.
One can accordingly compute two types of density current:
the canonical density currents
\begin{multline}
\boldsymbol{J}^C=
\mathrm{Re}\left[
\psi^*(x,y,t)
\left(-i\frac{\hbar}{m}\nabla\right)
\psi(x,y,t)
\right]\\
=\mathrm{Re}\left[
\psi^*(x,y,t)
\left\langle x,y\left\vert\frac{\hat{\boldsymbol{p}}}{m}\right\vert \psi \right\rangle
\right],\label{denscurr:C}
\end{multline}
and the kinetic density current
\begin{multline}
\boldsymbol{J}^K
=\mathrm{Re}\left[
\psi^*(x,y,t)
\left(-i\frac{\hbar}{m}\nabla+\frac{e}{m}\boldsymbol{A}\right)
\psi(x,y,t)
\right]
\\
=\mathrm{Re}\left[
\psi^*(x,y,t)
\left\langle x,y\left\vert\hat{\boldsymbol{v}}\right\vert \psi \right\rangle
\right],\label{denscurr:K}
\end{multline}
where $\hat{\boldsymbol{v}}=(\hat{\boldsymbol{p}}+e\hat{\boldsymbol{A}})/m$ is
the velocity operator.
These currents, specially the kinetic one,  are of great experimental significance.
The kinetic density current is associated to the Bohmian velocity
by $\boldsymbol{v}^K=\boldsymbol{J}^K/\rho$. In turn, the Bohmian trajectories
arising from the  $\boldsymbol{v}^K$ streamlines can be measured
by averaging over large numbers of identical
single-particle events \cite{Kocsis1170,1367-2630-15-7-073022,ISI:000281824900033}.

The relation that $S$ holds with the kinetic and canonical density current
is of particular interest to the present work.
It can be easily shown that the topological charge may
be associated to the canonical density current by
\begin{equation}
S=\frac{m}{2\pi\hbar}\oint_{\mathcal{C}} \boldsymbol{v}^C\cdot d\boldsymbol{l},
\end{equation} 
where the velocity $\boldsymbol{v}^C=\boldsymbol{J}^C/\rho$ is the
canonical version of the Bohmian velocity.
As the canonical and kinetic currents only differ by a term proportional
to the vector potential, the Bohmian velocity is
also connected  to the topological charge through the 
relation
\begin{equation}
S=\frac{m}{2\pi\hbar}\oint_{\mathcal{C}} \boldsymbol{v}^K\cdot d\boldsymbol{l}
-\frac{e}{2\pi \hbar}\Phi_B.\label{topo:kinetic}
\end{equation}
In the previous expression, $\Phi_B$ is the magnetic flux passing through
 the surface enclosed by $\mathcal{C}$.
If $\mathcal{C}$ is very close to the singularity
$\Phi_B\approx 0$ and the topological charge
can be estimated entirely from the closed integral
of $\boldsymbol{v}^K$.

Close to a singularity of the form (\ref{singu:form})
the canonical current components take the form
\begin{eqnarray}
J^C_x &=& -S\left\vert\gamma\beta\right\vert \left\vert a\right\vert^2\frac{\hbar}{m}
\left\vert \bar z-\bar z_j\right\vert^{2(s-1)}(y-y_j),
\label{singu:C:Jx}\\
J^C_y &=& S\left\vert\gamma\beta\right\vert \left\vert a\right\vert^2\frac{\hbar}{m}
\left\vert \bar z-\bar z_j\right\vert^{2(s-1)}(x-x_j).
\label{singu:C:Jy}
\end{eqnarray}
From these expressions, it is clear that
the canonical current close to the singularity
spins in the direction indicated by the topological charge.
Therefore, positive and negative topological charges yield
counterclockwise and clockwise currents respectively whereas
a vanishing topological charge yields a vanishing canonical current.
Moreover, the topological charge can be determined from the
shape of the Bohmian stream lines.
By inserting the vector potential (\ref{sym:gauge}) in
(\ref{denscurr:C})
the kinetic current components yield
\begin{eqnarray}
J^K_x &=& -\frac{\left\vert a\right\vert^2}{m}
\left\vert \bar z-\bar z_j\right\vert^{2(s-1)}\nonumber\\
&&\times
\left[S\hbar\left\vert\gamma\beta\right\vert(y-y_j)+\left\vert \bar z-\bar z_j\right\vert^2
\frac{eB y}{2}
\right],\label{singu:K:Jx}\\
J^K_y &=& \frac{\left\vert a\right\vert^2}{m}
\left\vert \bar z-\bar z_j\right\vert^{2(s-1)}\nonumber\\
&&\times
\left[S\hbar\left\vert\gamma\beta\right\vert(x-x_j)
+\left\vert \bar z-\bar z_j\right\vert^2
\frac{eB x}{2}
\right].\label{singu:K:Jy}
\end{eqnarray}
The components of the Bohmian velocity near the singularity
are therefore
\begin{eqnarray}
v^K_x &=& -\frac{1}{m \left\vert \bar z-\bar z_j\right\vert^{2}}
\nonumber\\
&&\times
\left[S\hbar\left\vert\gamma\beta\right\vert(y-y_j)+\left\vert \bar z-\bar z_j\right\vert^2
\frac{eB y}{2}
\right],\label{singu:K:vx}\\
v^K_y &=& \frac{1}{m \left\vert \bar z-\bar z_j\right\vert^{2}}
\nonumber\\
&&\times
\left[S\hbar\left\vert\gamma\beta\right\vert(x-x_j)
+\left\vert \bar z-\bar z_j\right\vert^2
\frac{eB x}{2}
\right].\label{singu:K:vy}
\end{eqnarray}
The last four equations show that close to the singularity the
direction of the kinetic density current is also determined by the
the sign of the topological charge. Nevertheless the
kinetic current might change direction away from
the singularity when $S<0$.
This is a characteristic feature of singularities having
the opposite topological charge sign compared to the classical
angular momentum of the electron.
As we discuss below, this is exactly
true for LG beams.
It is worthwhile to notice that close to a singularity 
($ \left\vert \bar z-\bar z_j\right\vert\approx 0$) the dominant contribution
comes from the term proportional to $S$.
Therefore
the kinetic and the canonical currents as well as the
velocities are similar, i.e. $\boldsymbol{J}^K\approx \boldsymbol{J}^C$ and
\begin{eqnarray}
v^K_x \approx v^C_x &=&
 -\frac{S\hbar\left\vert\gamma\beta\right\vert}{m \left\vert \bar z-\bar z_j\right\vert^{2}}
 (y-y_j),\label{vBx:app}\\
v^K_y \approx v^C_y &=& \frac{S\hbar\left\vert\gamma\beta\right\vert}{m \left\vert \bar z-\bar z_j\right\vert^{2}}
(x-x_j).\label{vBy:app}
\end{eqnarray}

In order to prove these results for the particular
case of BEC's and HG beams
we express the canonical and kinetic currents
in terms of the LG beams.
Calculating the explicit form of the velocity operators
in terms of the rising and lowering operators
(\ref{op:b})-(\ref{op:ct}) we obtain
\begin{eqnarray}
\hat v_x &=& \frac{\partial \hat H}{\partial \hat p_x}
=\frac{1}{m}\left(\hat p_x-\frac{m\omega}{2}\hat y\right)
=\frac{i\omega l_B}{\sqrt{2}}\left(\hat c^\dag-\hat c\right),
\label{vx:updown}\\
\hat v_y &=& \frac{\partial \hat H}{\partial \hat p_y}
=\frac{1}{m}\left(\hat p_y+\frac{m\omega}{2}\hat x\right)
=\frac{\omega l_B}{\sqrt{2}}\left(\hat c^\dag+\hat c\right).
\label{vy:updown}
\end{eqnarray}
Similarly, for the momentum we have
\begin{eqnarray}
\hat p_x &=& i \frac{m\omega l_B}{2\sqrt{2}}\left(\hat c^\dag-\hat c+\hat b^\dag-\hat b\right),
\label{px:updown}\\
\hat p_y &=&  \frac{m\omega l_B}{2\sqrt{2}}\left(\hat c^\dag+\hat c-\hat b^\dag-\hat b\right).
\label{py:updown}
\end{eqnarray}

In order to get an insight on the behaviour of these
currents in a simple case let us first calculate
them for pure LG beams.
Using the generalized Laguerre polynomials' recurrence relation
\begin{equation}
L_{k+1}^{j-1}(x)=\frac{j-x}{k+1}L_{k}^{j}(x)
-\frac{x}{k+1}L_{k-1}^{j+1}(x),
\end{equation}
we obtain the kinetic density currents components for LG beams
\begin{eqnarray}
J_x^{LG,K} &=&\mathrm{Re}\left[\frac{i\omega l_B}{\sqrt{2}}\psi^{LG*}_{l,n}(x,y)
\left\langle x,y\left\vert
\hat c^\dag-\hat c\right\vert l,n\right\rangle\right]\nonumber\\
&&=
-\rho^{LG}_{l,n}\left(x,y\right)
\frac{\omega l_B^2}{r}\left(l-n+\frac{r^2}{2l_B^2}\right)
\sin\theta,\\
J_y^{LG,K} &=&
\mathrm{Re}\left[\frac{\omega l_B}{\sqrt{2}}\psi^{LG*}_{l,n}(x,y)
\left\langle x,y\left\vert\hat c^\dag+\hat c
\right\vert l,n\right\rangle\right]\nonumber\\
&&=\rho^{LG}_{l,n}\left(x,y\right)
\frac{\omega l_B^2}{r}\left(l-n+\frac{r^2}{2l_B^2}\right)
\cos\theta,
\end{eqnarray}
where $r=\sqrt{x^2+y^2}$ and $\theta=\arctan\left(y/x\right)$.
These relations are consistent with the results presented in
Ref. \cite{PhysRevX.2.041011}.
Replacing  $\hat p_x/m=\hat v_x+\omega \hat y/2$ and
$\hat p_y/m=\hat v_y-\omega \hat x/2$
into the kinetic density current components, the canonical
ones are readily obtained 
\begin{eqnarray}
J_x^{LG,C} &=&\mathrm{Re}\bigg[\psi_{l,n}^*(x,y)\nonumber \\
&&\left. \times \left\langle x,y\left\vert
i\frac{m\omega l_B}{\sqrt{2}}\left(\hat c^\dag-\hat c\right)
+\frac{\omega}{2}\hat y\right\vert l,n\right\rangle\right]\nonumber \\
&&=-\rho^{LG}_{l,n}\left(r\right)
\frac{\omega l_B^2}{r}\left(l-n\right)
\sin\theta
,\\
J_y^{LG,C} &=&
\mathrm{Re}\bigg[\psi_{l,n}^*(x,y)\nonumber \\
&&\times\left.\left\langle x,y\left\vert \frac{m\omega l_B}{\sqrt{2}}\left(\hat c^\dag+\hat c\right)
-\frac{\omega}{2}\hat x
\right\vert l,n\right\rangle\right]\nonumber \\
&&=\rho^{LG}_{l,n}\left(r\right)
\frac{\omega l_B^2}{r}\left(l-n\right)
\cos\theta .
\end{eqnarray}
These equations clearly show that both kinetic and canonical density currents
similarly circulate around the singularity
located at $z=0$. They spin in clockwise or counterclockwise
direction depending on the sign of the angular momentum or topological charge $S=l-n$.
This is a confirmation
of the rather general results obtained in
Eqs. (\ref{singu:C:Jx})-(\ref{singu:K:Jy}).
Indeed, it has been shown that for $m=-1$, the canonical density current spins
around the singularity in counterclockwise direction, for $m=1$ it spins in
clockwise direction and for $m=0$ the canonical current vanishes\cite{PhysRevX.4.011013}.
The kinetic density current, however, exhibits a more complicated behaviour.
Whereas the canonical density current  spins only in the direction
indicated by the topological charge $S$,
the kinetic density current changes direction in the orbit
$r=l_B\sqrt{2(n-l)}$ provided that the angular momentum is negative ($l-n<0$).
This orbit corresponds precisely to the classical trajectory of the electron.

All of these features are clearly illustrated by plotting
the path integrals
\begin{eqnarray}
\eta^K &=& \oint_{\mathcal{C}}\boldsymbol{J^K}\cdot d\boldsymbol{l},\label{etaK}\\
\eta^C &=& \oint_{\mathcal{C}}\boldsymbol{J^C}\cdot d\boldsymbol{l},\label{etaC}
\end{eqnarray}
where $\mathcal{C}$ is a circle of radius $r$.
These two gauge  invariant functions differ only by a multiple of the Dirac phase.

The gain of using these two functions becomes evident
when we calculate them in the proximity of a
singularity of the type (\ref{singu:form}) obtaining
\begin{eqnarray}
\eta^C &=& 2\pi \omega l_B^2 S \rho(r),\label{etac:singu}\\
\eta^K &=& 2\pi \omega l_B^2
\left(S +\frac{\gamma+\beta}{2\gamma\beta}
\frac{r^2}{2l_B^2}\right)
\rho(r),\label{etak:singu}
\end{eqnarray}
where $\rho(r)=\left\vert a\right\vert^2 r^{2s}$ is the probability
density close to the singularity $z_j$.
These equations imply that the topological charge
can be computed analytically by taking the
limit as $z$ approaches $z_j$ of 
either $\eta^K/2\pi \omega l_B^2\rho$ or
$\eta^C/2\pi \omega l_B^2\rho$. The functions $\eta^K$ and $\eta^C$ 
are easier to calculate than the actual
phase change around a singularity $\int d\varphi$
that in many occasions can only be computed numerically.

The particular case of LG beams is quite illustrative
in this matter.
LG beams yield
\begin{eqnarray}
\eta^K &=& 2\pi\omega l_B^2\left(l-n+\frac{r^2}{2l_B^2}\right)\rho^{LG}_{l,n}\left(r\right),\\
\eta^C &=& 2\pi\omega l_B^2\left(l-n\right)\rho^{LG}_{l,n}\left(r\right).
\end{eqnarray}
Notice that (\ref{etac:singu}) and (\ref{etak:singu}) reduce to
the expressions above provided that we set $S=l-n$
and the integral is performed
around a circle, namely $\gamma=\beta=1$.

From the previous equations we observe that while
$\eta^K>0$ for $m=0,1$ for all $r$ values,
$\eta^K>0$ with $m=-1$ presents a sign change
at $r=\sqrt{2}l_B$ characteristic of diamagnetic states.
The expression of $\eta^C$ clearly shows that the canonical currents
spin in the expected direction given by the sign of the
angular momentum and the topological charge $S$.
By taking the limit $r\rightarrow 0$ of $\eta^C/2\pi \omega l_B^2\rho$
or $\eta^K/2\pi \omega l_B^2\rho$ we obtain the topological charge $S$
confirming the result in Eqs. (\ref{etac:singu}) and (\ref{etak:singu}).

By inserting the velocity operators  (\ref{vx:updown}) and (\ref{vy:updown})
into  the definition of the density current (\ref{denscurr:K})
for a general state of the form (\ref{psi:gen}) we get
\begin{eqnarray}
J_x^{K}=&&\mathrm{Re}
\bigg[i\frac{\omega l_B}{\sqrt{2}}
\sum_{l,l^\prime,n,n^\prime}A^*_{l^\prime,n^\prime}A_{l,n}
\mathrm{e}^{i\omega(l^\prime -l)t}\psi_{l^\prime,n^\prime}^{LG*}\nonumber\\
&&\times \left(
\sqrt{l+1}\psi^{LG}_{l+1,n}-\sqrt{l}\,\,\psi^{LG}_{l-1,n}\right)\bigg],
\label{jxK:gen}\\
J_y^{K}=&&\mathrm{Re}
\bigg[\frac{\omega l_B}{\sqrt{2}}
\sum_{q,p=q_{min}}^{q_{max}}A^{j,k}_pA^{j,k}_q
\mathrm{e}^{i\omega(l^\prime -l)t}\psi_{l^\prime,n^\prime}^{LG*}\nonumber\\
&&\times \left(
\sqrt{l+1}\psi^{LG}_{l+1,n}+\sqrt{l}\,\,\psi^{LG}_{l-1,n}\right)\bigg].
\label{jyK:gen}
\end{eqnarray}
Similarly, inserting the components of the
momentum operator (\ref{px:updown})-(\ref{py:updown})
into (\ref{denscurr:C}) the canonical current is readily
obtained as
\begin{eqnarray}
J_x^{C}&=&\mathrm{Re}
\bigg[i\frac{\omega l_B}{2\sqrt{2}}
\sum_{l,l^\prime,n,n^\prime}A^*_{l^\prime,n^\prime}A_{l,n}
\mathrm{e}^{i\omega(l^\prime -l)t}\psi_{l^\prime,n^\prime}^{LG*}\nonumber\\
&&\times \big(
\sqrt{l+1}\psi^{LG}_{l+1,n}-\sqrt{l}\,\,\psi^{LG}_{l-1,n}\nonumber\\
&&+\sqrt{n+1}\psi^{LG}_{l,n+1}-\sqrt{n}\,\,\psi^{LG}_{l,n-1}\big)\bigg],\label{jxC:gen}\\
J_y^{C}&=&\mathrm{Re}
\bigg[\frac{\omega l_B}{2\sqrt{2}}
\sum_{l,l^\prime,n,n^\prime}A^*_{l^\prime,n^\prime}A_{l,n}
\mathrm{e}^{i\omega(l^\prime -l)t}\psi_{l^\prime,n^\prime}^{LG*}\nonumber\\
&&\times \big(
\sqrt{l+1}\psi^{LG}_{l+1,n}+\sqrt{l}\,\,\psi^{LG}_{l-1,n}\nonumber\\
&&-\sqrt{n+1}\psi^{LG}_{l,n+1}-\sqrt{n}\,\,\psi^{LG}_{l,n-1}\big)\bigg].\label{jyC:gen}
\end{eqnarray}

In the particular case of
HG beams the kinetic density current
is given by
\begin{eqnarray}
J_x^{HG,K}&&=\mathrm{Re}
\bigg[i\frac{\omega l_B}{\sqrt{2}}
\sum_{q,p=q_{min}}^{q_{max}}\left(A^{j,k}_p\right)^*A^{j,k}_q
\mathrm{e}^{i\omega(p-q)t}\psi_{p,j+k-p}^*\nonumber\\
&&\times \left(
\sqrt{q+1}\psi_{q+1,j+k-q}-\sqrt{q}\psi_{q-1,j+k-q}\right)\bigg],
\label{jxHGK}\\
J_y^{HG,K}&&=\mathrm{Re}
\bigg[\frac{\omega l_B}{\sqrt{2}}
\sum_{q,p=q_{min}}^{q_{max}}\left(A^{j,k}_p\right)^{*} A^{j,k}_q
\mathrm{e}^{i\omega(p-q)t}\psi_{p,j+k-p}^*\nonumber\\
&&\times \left(
\sqrt{q+1}\psi_{q+1,j+k-q}+\sqrt{q}\psi_{q-1,j+k-q}\right)\bigg].
\label{jyHGK}
\end{eqnarray}
For the canonical density current of HG beams we have
\begin{eqnarray}
J_x^{HG,C}=&&\mathrm{Re}
\bigg[i\frac{\omega l_B}{2\sqrt{2}}
\sum_{q,p=q_{min}}^{q_{max}}\left(A^{j,k}_p\right)^* A^{j,k}_q
\mathrm{e}^{i\omega(p-q)t}\psi_{p,j+k-p}^*\nonumber\\
&&\times \big(
\sqrt{q+1}\psi_{q+1,j+k-q}-\sqrt{q}\psi_{q-1,j+k-q}\nonumber\\
&&+\sqrt{j+k-q+1}\psi_{q,j+k-q+1}\nonumber\\
&&-\sqrt{j+k-q-1}\psi_{q,j+k-q-1}\big)\bigg],\label{jxHGC}\\
J_y^{HG,C}&=&\mathrm{Re}
\bigg[\frac{\omega l_B}{2\sqrt{2}}
\sum_{q,p=q_{min}}^{q_{max}}\left(A^{j,k}_p\right)^* A^{j,k}_q
\mathrm{e}^{i\omega(p-q)t}\psi_{p,j+k-p}^*\nonumber\\
&&\times \big(
\sqrt{q+1}\psi_{q+1,j+k-q}+\sqrt{q}\psi_{q-1,j+k-q}\nonumber\\
&&-\sqrt{j+k-q+1}\psi_{q,j+k-q+1}\nonumber\\
&&-\sqrt{j+k-q-1}\psi_{q,j+k-q-1}\big)\bigg].\label{jyHGC}
\end{eqnarray}
For standard HG beam, $q_{min}=0$ and
$q_{max}=j+k$.
Instead, for the positive and negative part of HG
beams $q_{min}$ and $q_{max}$ are given by
Eqs. (\ref{q:m:min})-(\ref{q:p:max}).
In the following sections, the expressions above will be very useful
in exploring the vorticity of density currents  around singularities.

\section{Results and Discussion}\label{res:disc}

In this section we wish to prove the general
relations between the singularities' mathematical
structure and  the kinetic and canonical currents in more
involved electron beams.
In order to analyse sufficiently complex systems
with a rich singularity structure
we have chosen HG and BEC beams.

We begin by investigating $\psi^{HG}_{3,3}$
and $\psi^{\pm HG}_{3,3}$.
The state $\psi^{HG}_{3,3}$ has the structure of a balanced beam
given by
\begin{multline}
\psi^{HG}_{3,3}=-\frac{i}{4} \left( \sqrt{5} \psi^{LG} _{0,6}
- \sqrt{3} \psi^{LG} _{2,4}\right.\\
\left.+ \sqrt{3} \psi^{LG} _{4,2}
- \sqrt{5} \psi^{LG}_{6,0}\right).
\end{multline}
Substituting the explicit form of $\psi^{HG}_{3,3}$
into Eqs. (\ref{jxHGC}) and (\ref{jyHGC})
the canonical density current vanishes
$\boldsymbol{J}_{3,3}^{HG,C}=\boldsymbol{0}$ in the
whole domain of the wave function.
Thereby, the kinetic density current $\boldsymbol{J}^{HG,K}_{3,3}=eB(-y\boldsymbol{i}
+x\boldsymbol{j})\rho_{3,3}^{HG}/m$, that only contains
the terms arising from the vector potential,
spins in the counterclockwise direction as it can be seen
in Fig. \ref{figure1}.
In contrast to LG beams, that have singularities
in isolated points, HG beams present singular points
arranged as lines \cite{Gbur2003117}.
In Fig. \ref{figure1} (a) we show the structure of the
$\psi^{HG}_{3,3}$ singularities. They can conveniently
be grouped in one point [(purple) dot] and 6 lines (dashed lines)
listed in rows 3-7 of Table \ref{singu:table}.
Except for the point (row 3) all of these singularities
can be put in the form (\ref{singu:form}) where $\gamma\beta=0$.
Under the arguments presented in Secs. \ref{singu:struc}
and \ref{currentsKC} the canonical currents close to these singularities
must vanish as expected.
Figs. \ref{figure1} (a), (b) and (c) also show the time evolution of the
probability density $\rho^{HG}_{3,3}$ for times $t=0$, $t=\pi/4\omega$ and
$t=\pi/2\omega$ respectively. Given that the beam is a superposition
of states with different eigenenergies it presents a nontrivial time evolution.
In this case, the singularity lines rotate in the counterclockwise direction,
in the same direction a classical electron would move under the action of
a perpendicular magnetic field.
This behaviour is observed in all of the singularities
presented from here on.

The singularities in states $\psi^{\pm HG}_{3,3}$
have a completely different structure.
They can be expressed in terms of the LG states as
\begin{eqnarray}
\psi^{-HG}_{3,3} &=& -\frac{i}{4} \left( \sqrt{5} \psi^{LG} _{0,6}
- \sqrt{3} \psi^{LG} _{2,4}\right), \\
\psi^{+HG}_{3,3} &=&-\frac{i}{4} \left( \sqrt{3} \psi^{LG} _{4,2}
- \sqrt{5} \psi^{LG}_{6,0}\right).
\end{eqnarray}
The above states have  a total of 13 singularities
arranged as isolated points. These are shown in Figs. \ref{figure2} (a)
and \ref{figure3} (a) as dots and are listed in rows 8-15 in Table \ref{singu:table}.
A more detailed diagram of the four different types of
singularities and their corresponding topological charges is
shown in Fig. \ref{figure4}.  The singularity located at the center
$z=0$ has a topological charge $S^{\pm}_{3,3}=\pm 2$.
Along the lines forming angles of $0^\circ$, $90^\circ$, $180^\circ$ and $270^\circ$
we have two different types of singularities. The first type with $S^{\pm}_{3,3}=\pm 1$
and the second with $S^{\pm}_{3,3}=\mp 1$. Along lines forming angles of $45^\circ$, $135^\circ$, $225^\circ$
and $315^\circ$ with the $x$ axis we find the fourth kind of singularity having
a vanishing topological charge $S^{\pm}_{3,3}=0$.
By using Eqs.  (\ref{jxHGK}), (\ref{jyHGK}), (\ref{jxHGC}) and (\ref{jyHGC})
we can compute the components of the kinetic and canonical density currents.
The  streamlines of the $\psi^{+HG}_{3,3}$ kinetic density current, namely
the Bohmian trajectories, are presented in 
Figs. \ref{figure2} (a) and (b) while (c) and (d) show
the canonical density current.
Similarly, Figs. \ref{figure3} (a) and (b) plot the kinetic density current
and (c) and (d) the canonical one for $\psi^{-HG}_{3,3}$.
We immediately notice that, in contrast to $\psi^{+HG}_{3,3}$, the streamlines in
$\psi^{-HG}_{3,3}$ switch from clockwise to counterclockwise direction
as we move away from the center of the beam.
As it was discussed for LG beams, it is a distinctive feature
of singularities having the opposite topological charge sign to
the spinning direction
of the classical electron.

The same vector fields close to the four types
of singularities
are shown in Figs. \ref{figure5} and
\ref{figure6} for $\psi^{+HG}_{3,3}$ and
$\psi^{-HG}_{3,3}$ respectively.
Only the canonical current is shown in this case
because, given the proximity to the singularities,
the terms related to the vector potential do not strongly distort
the general form of the current streamlines.
In each case we confirm that the canonical current spins in the direction
indicated by the topological charge.

Figs. \ref{figure7} (a) and (b) show plots of $\eta^K$ and $\eta^C$ respectively.
In this case they are calculated through
Eqs. (\ref{etaK}) and (\ref{etaC}) integrating
around a circle of radius $r=\sqrt{x^2+y^2}$.
In these figures it is possible to note
that the kinetic density current of the negative angular momentum
part of $\psi^{HG}_{3,3}$ presents a sign
change that is consistent with the relative sign of the
singularity's topological charge
compared to the angular momentum of the classical electron.
This same feature was discussed above regarding the Bohmian trajectories
observed in Figs. \ref{figure3} and \ref{figure4}.

Let us now turn our attention to the calculation
of the topological charge via the limit
$S=\lim_{r \rightarrow 0}\eta^K/2\pi \omega l_B^2\rho
=\lim_{r \rightarrow 0}\eta^C/2\pi \omega l_B^2\rho$.
In Fig. \ref{figure8} we have plotted
$\eta^K/2\pi \omega l_B^2\rho$ and $\eta^C/2\pi \omega l_B^2\rho$
as functions of the distance to the singularity $r$ for the
four types of singularities (shown in Figs. \ref{figure5} and \ref{figure6})
found in the $\psi^{\pm HG}_{3,3}$ beams.
Figs. \ref{figure8} (a), (b), (c) and (d) correspond to the singularities
listed in Table \ref{singu:table} in rows 8-11 for
$\psi^{-HG}_{3,3}$ and rows 12-15 for $\psi^{+HG}_{3,3}$.
Given the high symmetry of $S^{\pm HG}_{3,3}=\pm 2$
[Fig. \ref{figure8} (a)], the corresponding
$\eta^K/2\pi \omega l_B^2\rho$ and $\eta^C/2\pi \omega l_B^2\rho$
where calculated exactly.
However, in order to plot $\eta^K/2\pi \omega l_B^2\rho$
and $\eta^C/2\pi \omega l_B^2\rho$ for the three remaining
singularities $S^{\pm HG}_{3,3}=\pm 1,\mp 1,0$ with lower symmetry,
a Taylor expansion of  order 12
in $r$ around the singularity ($r=0$) was used.
Even though such high order expansions are needed
to obtain accurate plots, the calculation of the topological
charge only requires a Taylor expansion of the lowest non-vanishing order.
For example, expanding $\eta^C/2\pi \omega l_B^2\rho$
to the first order in $r$
for $S^{-HG}_{3,3}=-1$ around $z=2l_B\sqrt{3/(2+\sqrt{2})}+i 0$
with $\gamma=1$ and $\beta=(\sqrt{2}-1)/2$
(see Table \ref{singu:table}, row 9) we get
\begin{multline}
\eta^C/2\pi \omega l_B^2\rho \approx
-1+\frac{r}{12 l_B}
\sqrt{3 \left(2+\sqrt{2}\right)}
\cos\theta\\
\times\left[10 \sqrt{2}-3-\left(7+2 \sqrt{2}\right) \cos 2\theta\right].
\end{multline}
Similarly for $S^{-HG}_{3,3}=1$, expanding around
$z=2l_B\sqrt{3/(2-\sqrt{2})}+i 0$ with $\gamma=1$ and $\beta=(1+\sqrt{2})/2$
(see Table \ref{singu:table}, row 10) we have
\begin{multline}
\eta^C/2\pi \omega l_B^2\rho \approx 1+
\frac{r}{24 l_B}\sqrt{3 \left(2+\sqrt{2}\right)} \\
 \times\left[  \left(7+2 \sqrt{2}\right) \cos 3 \theta\right.
 \left.-\left(18 \sqrt{2}-13\right) \cos \theta  \right],
\end{multline}
and for $S^{-HG}_{3,3}=0$, expanding around
$z=\sqrt{3}+i\sqrt{3}$ with $\gamma=1$ and $\beta=0$
(see Table \ref{singu:table}, row 10) we have
\begin{equation}
\eta^C/2\pi \omega l_B^2\rho=0.
\end{equation}
The three functions clearly reduce to their corresponding
topological charges when the limit $r\rightarrow 0$ is taken.

To further prove the formulas obtained in Sec. \ref{currentsKC} we also study 
BEC states. These
have $n$ singularities arranged as isolated and equally spaced points.
In Fig. \ref{figure9} we illustrate the time evolution of the $\psi^{+BEC}_3$ beam.
The vector field represents the $\boldsymbol{J}^K$ density currents in (a) and (b)
and $\boldsymbol{J}^C$ in (c) and (d). Fig. \ref{figure10} shows the time evolution of the
$\psi^{-BEC}_3$ beam. We observe again the features seen for LG and HG
beams: i) in the positive beam  $\psi^{+BEC}_3$ both kinetic and canonical currents 
spin in counterclockwise direction, ii) in the negative beam $\psi^{-BEC}_3$
the kinetic current near the singularities spin in the clockwise direction and iii) as we move
away from the singularities the counterclockwise direction of the kinetic current
is recovered. The positive beam $\psi^{+BEC}_3$ is not an energy eigenstate
and therefore it rotates in the counterclockwise direction as it evolves in time.
The negative beam, however, does not rotate since it is a stationary state,
composed of two LG beams with the same energy $\hbar \omega /2$ ($l=0$). 

The behaviour of the kinetic currents close to the singularities is shown
in Fig. \ref{figure11} for (a) $\psi^{+BEC}_3$ and
(b) $\psi^{-BEC}_3$  (see row 16 in Table \ref{singu:table} ) .
We observe that the direction of the kinetic current is determined again
by the sign of the topological charge. 

Now, expanding $\eta^C/2\pi \omega l_B^2\rho$ and $\eta^K/2\pi \omega l_B^2\rho$
to the first order, we obtain the $\psi^{\pm BEC}_3$ topological charges.
For $S^{-BEC}_{3}=-1$ we have $\gamma=1$ and $\beta=-1$
(see row 16 in Table \ref{singu:table})
\begin{multline}
\eta^C/2\pi \omega l_B^2\rho \approx 
\eta^K/2\pi \omega l_B^2\rho \approx
 -1+\frac{r}{\sqrt{3} l_B}\left(\frac{3}{2}\right)^{1/3}\\
\times\left[1-\left(\frac{3}{2}\right)^{1/3}\right]
 \left(\frac{1}{\sqrt{3}}\cos\theta+\sin\theta\right).
\end{multline}
Similarly the singularity $S^{+BEC}_{3}=1$ we have $\gamma=1$ and $\beta=1$
(see row 17 in Table \ref{singu:table})
\begin{multline}
\eta^C/2\pi \omega l_B^2\rho \approx 
\eta^K/2\pi \omega l_B^2\rho \approx
 1-\frac{r}{\sqrt{3} l_B}\left(\frac{3}{2}\right)^{1/3}\\
\times \left[1-\left(\frac{3}{2}\right)^{1/3}\right]
 \left(\frac{1}{\sqrt{3}}\cos\theta+\sin\theta\right).
\end{multline}
Again we observe that the limit of the expressions above as the distance
to the singularity approaches zero is equal to the topological charge.

\section{Finding the mathematical structure
of a singularity from the Bohmian streamlines}
\label{outline}

Using the previous relations,
in this section we outline a method to obtain the
mathematical structure of the wave function near a singularity.
Starting from the Bohmian streamlines,  we work backward
obtaining the main parameters that define the wave function.

As an example, let us consider the beam characterized by the wave function
\begin{multline}
\psi(x,y)=\frac{1}{\mathcal{N}l_B^4}\exp\left(-\frac{x^2+y^2}{4l_B^2}\right)\\
\times \Bigg\{
\left[
\frac{\sqrt{3}}{2}\left(x-3l_B\right)
+\frac{1}{2}\left(y-l_B\right)
\right]\\
+i\frac{3}{2}\left[
\frac{\sqrt{3}}{2}\left(y-l_B\right)
-\frac{1}{2}\left(x-3l_B\right)
\right]
\Bigg\}^3,\label{exp:psi}
\end{multline}
where $\mathcal{N}=\sqrt{(6192176-1960965 \sqrt{3})\pi }/16$
is a normalisation constant.
From the structure of the wave function it is easy to show that
it only has one singularity located in $x_j=3l_B$ and $y_j=l_b$,
with $S=+3$, $\gamma=1$, $\beta=3/2$
and it is tilted $30^{\circ}$ with respect to the $x$-axis.
The wave function (\ref{exp:psi}) can be expressed in terms
of the LG modes as
\begin{multline}
\psi=
   A_{0,0}\psi^{LG}_{0,0}+
   A_{0,1}\psi^{LG}_{0,1}+
    A_{0,2}\psi^{LG}_{0,2}+
  A_{0,3}\psi^{LG}_{0,3}\\
  +A_{1,0} \psi^{LG}_{1,0}
  +A_{1,1}  \psi^{LG}_{1,1}
   +A_{1,2} \psi^{LG}_{1,2}
   +A_{2,0}  \psi^{LG}_{2,0}\\
   +A_{2,1}\psi^{LG}_{2,1}
   +A_{3,0}\psi^{LG}_{3,0},
\end{multline}
where the coefficients are given by
\begin{eqnarray}
A_{0,0} &=&\frac{1}{\mathcal{N}}
  \left(\frac{1}{4}-\frac{3 i}{16}\right)\nonumber\\
   &&\times\left[(-247-138 i)+(63+126 i)
   \sqrt{3}\right] \sqrt{\frac{\pi }{2}},\\
A_{0,1} &=& -\frac{1}{\mathcal{N}}\left(\frac{27}{32}+\frac{3 i}{16}\right)\nonumber\\
 &&\times \left[(5-18 i)+(6+7 i) \sqrt{3}\right] \sqrt{\pi },\\
A_{0,2} &=&   -\frac{1}{\mathcal{N}}\left(\frac{9}{32}-\frac{3 i}{32}\right)\nonumber\\
&&\times \left[(-3+2 i)+(4+3 i)
   \sqrt{3}\right] \sqrt{\pi }, \\
A_{0,3} &=&  -\frac{1}{8\mathcal{N}} i\sqrt{\frac{3 \pi }{2}},   \\
A_{1,0} &=&  -\frac{1}{\mathcal{N}}\left(\frac{15}{16}-\frac{135 i}{32}\right) \nonumber\\
   &&\times \left[(-18-13 i)+(1+6 i)   \sqrt{3}\right] \sqrt{\pi },\\
A_{1,1} &=& \frac{1}{\mathcal{N}} \left(\frac{15}{4}+\frac{1 5 i}{8}\right) 
   \left[3 \sqrt{3}-(1+4   i)\right]\sqrt{\frac{\pi }{2}} ,\\
A_{1,2} &=&  \frac{15}{16\mathcal{N}} \left(\sqrt{3}+i\right) \sqrt{\frac{\pi}{2}},\\
A_{2,0} &=&  \frac{1}{\mathcal{N}} \left(\frac{75}{32}+\frac{225 i}{32}\right)
    \left[(7+6 i)-i\sqrt{3}\right]\sqrt{\pi },\\
A_{2,1} &=& \frac{75}{16\mathcal{N}}  \left(\sqrt{3}-i\right) \sqrt{\frac{\pi }{2}} ,\\
A_{3,0} &=& -\frac{125}{8\mathcal{N}} i   \sqrt{\frac{3 \pi }{2}} .
\end{eqnarray}

\begin{table*}
\caption{List of singularities. The first column presents the state}\label{singu:table}
\begin{tabular}{ | l | l | l | l | l |}
\hline
 $\psi$ & $z_j$ & $S$ & $\lim_{z\rightarrow z_j}\psi$ & $\gamma\beta$ \\
\hline
\hline
\begin{tabular}{l}
 $\psi_{l,n}^{LG}$,\\
  $n>l$\\
  \end{tabular}
   &
  $0$ & $S^{LG}_{l,n}=l-n$ & $\frac{l!}{\pi n!  (2l_B^2)^{n-l}}L_l^{n-l}(0)\left[(x-x_j)-i(y-y_j)\right]^{n-l}$ & $\gamma\beta <0$ \\
\hline
\begin{tabular}{l}
 $\psi_{l,n}^{LG}$,\\
  $l>n$\\
  \end{tabular}
   &
  $0$ & $S^{LG}_{l,n}=l-n$ & $\frac{n!}{\pi l!  (2l_B^2)^{l-n}}L_n^{l-n}(0)\left[(x-x_j)+i(y-y_j)\right]^{l-n}$ & $\gamma\beta >0$ \\
\hline
\hline
 $\psi_{3,3}^{HG}$ &
  $0+i0$ &
  $S^{HG}_{3,3}=0$ &
  $\frac{3}{2l_B^3\sqrt{2\pi}}(x-x_j)(y-y_j)$ &
  $\gamma\beta=0$ \\
\hline
 $\psi_{3,3}^{HG}$ &
  $0+id, d\ne 0$ &
  $S^{HG}_{3,3}=0$ &
  $-\frac{d \left(d^2-3l_B^2\right) }{2 l_B^5\sqrt{2 \pi }}\mathrm{e}^{-\frac{d^2}{4l_B^2}}(x-x_j)$ &
  $\gamma\beta=0$ \\
\hline
 $\psi_{3,3}^{HG}$ &
  $d+i 0, d\ne 0$ &
  $S^{HG}_{3,3}=0$ &
  $-\frac{d \left(d^2-3l_B^2\right) }{2 l_B^5\sqrt{2 \pi }}\mathrm{e}^{-\frac{d^2}{4l_B^2}}(y-y_j)$ &
  $\gamma\beta=0$ \\
\hline
 $\psi_{3,3}^{HG}$ &
  $\pm \sqrt{3}l_B +id$ &
  $S^{HG}_{3,3}=0$ &
  $-\frac{d \left(d^2-3l_B^2\right)}{2 l_B^5\sqrt{2 \pi }}\mathrm{e}^{-\frac{d^2}{4l_B^2}-\frac{9}{4}}
  \left\{  (x-x_j)
  \mp 3\left[\frac{2l_B}{d} -\frac{d}{l_B}\left(1-\frac{4l_B^2}{d^2-3l_B^2}\right)\right] (y-y_j)
  \right\}$ &
  $\gamma\beta=0$ \\
\hline
 $\psi_{3,3}^{HG}$ &
  $ d \pm i\sqrt{3}l_B$ &
  $S^{HG}_{3,3}=0$ &
  $-\frac{d \left(d^2-3l_B^2\right)}{2 l_B^5\sqrt{2 \pi }}\mathrm{e}^{-\frac{d^2}{4l_B^2}-\frac{9}{4}}
  \left\{ \mp 3\left[\frac{2l_B}{d} -\frac{d}{l_B}\left(1-\frac{4l_B^2}{d^2-3l_B^2}\right)\right]  (x-x_j)
  (y-y_j)
  \right\}$ &
  $\gamma\beta=0$ \\
\hline
 $\psi_{3,3}^{-HG}$ &
  $0$ &
  $S^{-HG}_{3,3}=-2$ &
  $\frac{3 i}{8 \sqrt{2 \pi } l_B^3} \left[(x-x_j)-i (y-y_j)\right]^2$ &
  $\gamma\beta<0$ \\
\hline
 $\psi_{3,3}^{-HG}$ &
 \begin{tabular}{l}
  $2l_B\sqrt{\frac{3}{2+\sqrt{2}}} \mathrm{e}^{in\pi/4},$\\
  \,\,\,\,\,\,\, $ n=0,2,4,6$ 
  \end{tabular} &
  $S^{-HG}_{3,3}=-1$ &
  $-\frac{3}{2l_B^2} i \mathrm{e}^{\frac{3}{\sqrt{2}}-3} \sqrt{\frac{3 \left(10-7 \sqrt{2}\right)}{\pi }} \left[
  (x-x_j)  +\frac{i}{2}\left(1-\sqrt{2}\right)(y-y_j)\right]$ &
  $\gamma\beta<0$ \\
\hline
$\psi_{3,3}^{-HG}$ &
 \begin{tabular}{l}
  $2l_B\sqrt{\frac{3}{2-\sqrt{2}}} \mathrm{e}^{in\pi/4},$\\
  \,\,\,\,\,\,\,$n=0,2,4,6$
  \end{tabular} &
  $S^{-HG}_{3,3}=+1$ &
  $\frac{3}{2l_B^2} i \mathrm{e}^{-3-\frac{3}{\sqrt{2}}}
   \sqrt{\frac{3 \left(10+7 \sqrt{2}\right)}{\pi }}
   \left[(x-x_j)+\frac{i}{2}\left(1+\sqrt{2}\right)(y-y_j)\right]$ &
  $\gamma\beta>0$ \\
\hline
 $\psi_{3,3}^{-HG}$ &
 \begin{tabular}{l}
  $\sqrt{6}l_B\mathrm{e}^{i n\pi/4},$\\
   \,\,\,\,\,\,\,$n=1,3,5,7$
  \end{tabular} &
  $S^{-HG}_{3,3}=0$ &
  $\frac{3 i \sqrt{\frac{3}{2 \pi }}}{8 l_B^2\mathrm{e}^{3/2}} \left[(x-x_j)- (y-y_j)\right]$ &
  $\gamma\beta=0$ \\
\hline
$\psi_{3,3}^{+HG}$ &
  $0$ &
  $S^{+HG}_{3,3}=+2$ &
  $-\frac{3 i}{8l_B^3 \sqrt{2 \pi } } \left[(x-x_j)+i (y-y_j)\right]^2$ &
   \\
  \hline
 $\psi_{3,3}^{+HG}$ &
 \begin{tabular}{l}
  $2l_B\sqrt{\frac{3}{2+\sqrt{2}}}, \mathrm{e}^{in\pi/4},$\\
   \,\,\,\,\,\,\,$n=0,2,4,6$
   \end{tabular}&
  $S^{+HG}_{3,3}=+1$ &
  $\frac{3}{2l_B^2} i \mathrm{e}^{\frac{3}{\sqrt{2}}-3}
   \sqrt{\frac{3 \left(10-7 \sqrt{2}\right)}{\pi }}
   \left[(x-x_j)  -\frac{i}{2}\left(1-\sqrt{2}\right)(y-y_j)\right]$ &
  $\gamma\beta>0$ \\
\hline
 $\psi_{3,3}^{+HG}$ &
 \begin{tabular}{l}
  $2l_B\sqrt{\frac{3}{2-\sqrt{2}}} \mathrm{e}^{in\pi/4},$\\
   \,\,\,\,\,\,\,$n=0,2,4,6$
   \end{tabular} &
  $S^{+HG}_{3,3}=-1$ &
  $-\frac{3}{2l_B^2} i \mathrm{e}^{-3-\frac{3}{\sqrt{2}}}
 \sqrt{\frac{3 \left(10+7 \sqrt{2}\right)}{\pi }}\left[
  (x-x_j)-\frac{i}{2}\left(1+\sqrt{2}\right)(y-y_j)\right]$ &
  $\gamma\beta<0$ \\
\hline
$\psi_{3,3}^{+HG}$ &
\begin{tabular}{l}
  $\sqrt{6}l_B\mathrm{e}^{i n\pi/4},$\\
    \,\,\,\,\,\,\,$n=1,3,5,7$ 
    \end{tabular}&
  $S^{+HG}_{3,3}=0$ &
  $-\frac{3 i \sqrt{\frac{3}{2 \pi }}}{8l_B^2 \mathrm{e}^{3/2}} \left[(x-x_j)- (y-y_j)\right]$ &
  $\gamma\beta=0$ \\
\hline
\hline
$\psi_{3}^{-BEC}$ &
\begin{tabular}{l}
  $\sqrt[3]{2\sqrt{3}}l_B\mathrm{e}^{i n\pi/3},$\\
   \,\,\,\,\,\,\,$ n=1,3,5$ 
  \end{tabular}&
  $S^{-BEC}_{3}=-1$ &
  $-\frac{\sqrt[3]{3} \left(\sqrt{3}+3 i\right)
   \mathrm{e}^{-\frac{1}{2} \sqrt[3]{\frac{3}{2}}}}{ 2^{11/6}l_B^2
   \sqrt{5 \pi }} \left[(x-x_j)-i(y-y_j)\right]$ &
  $\gamma\beta<0$ \\  
\hline
$\psi_{3}^{+BEC}$ &
\begin{tabular}{l}
  $\sqrt[3]{2\sqrt{3}} l_B\mathrm{e}^{i n\pi/3},$\\
   \,\,\,\,\,\,\,$ n=1,3,5$
   \end{tabular} &
  $S^{+BEC}_{3}=+1$ &
  $-\frac{\sqrt[3]{3} \left(\sqrt{3}+3 i\right)
   \mathrm{e}^{-\frac{1}{2} \sqrt[3]{\frac{3}{2}}}}{ 2^{11/6}l_B^2
   \sqrt{5 \pi }} \left[(x-x_j)+i(y-y_j)\right]$ &
  $\gamma\beta>0$ \\
\hline
\end{tabular}
\end{table*}

\begin{figure}
\includegraphics[angle=0,width=0.5 \textwidth]{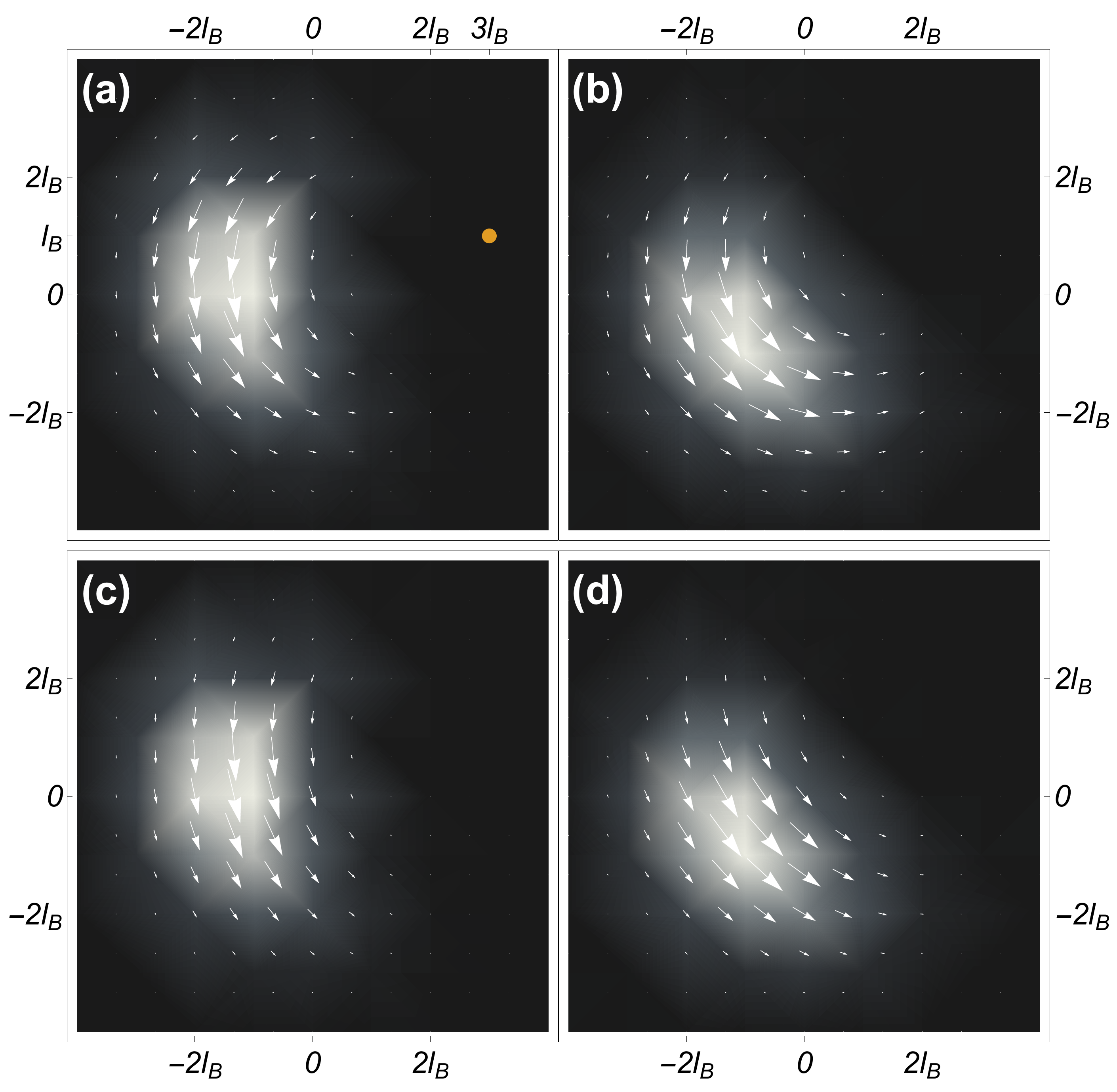}
\caption{Rotational dynamics of $\psi$.
Along the first row, in (a) and (b) we observe the time evolution of
$\rho=\left\vert \psi \right\vert^2$ and the vector field $\boldsymbol{J}^K$
for $t=0$ and $t=\pi/6\omega$ respectively.
In the second row, (c) and (d)  show
the time evolution of
$\rho$ and the vector field $\boldsymbol{J}^C$
for $t=0$ and $t=\pi/6\omega$ respectively.
The positions of the singularity $z_j=3l_B+il_B$ is
indicated with an (orange) dot in (a).}
\label{figure14}
\end{figure}

\begin{figure}
\includegraphics[angle=0,width=0.5 \textwidth]{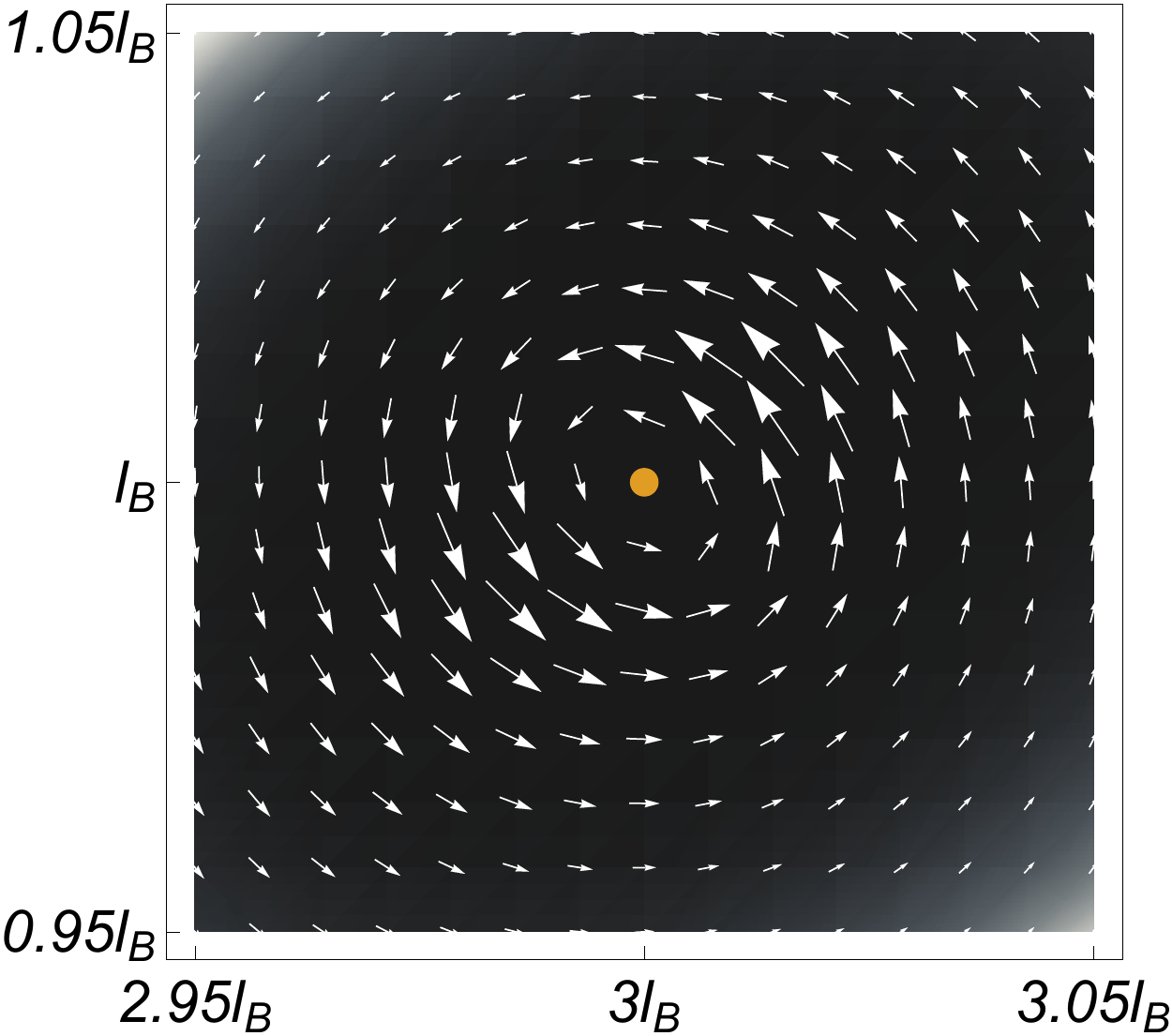}
\caption{ Bohmian velocity around the singularitiy $z_j=3l_B+il_B$ of $\psi$.
The position
of the singularity is indicated with a dot.
The density plot of the density probability $\rho=\left\vert \psi \right\vert^2$
and the corresponding Bohmian velocity field $\boldsymbol{v}^K$ are also shown.}
\label{figure15}
\end{figure}

\begin{figure}
\includegraphics[angle=0,width=0.5 \textwidth]{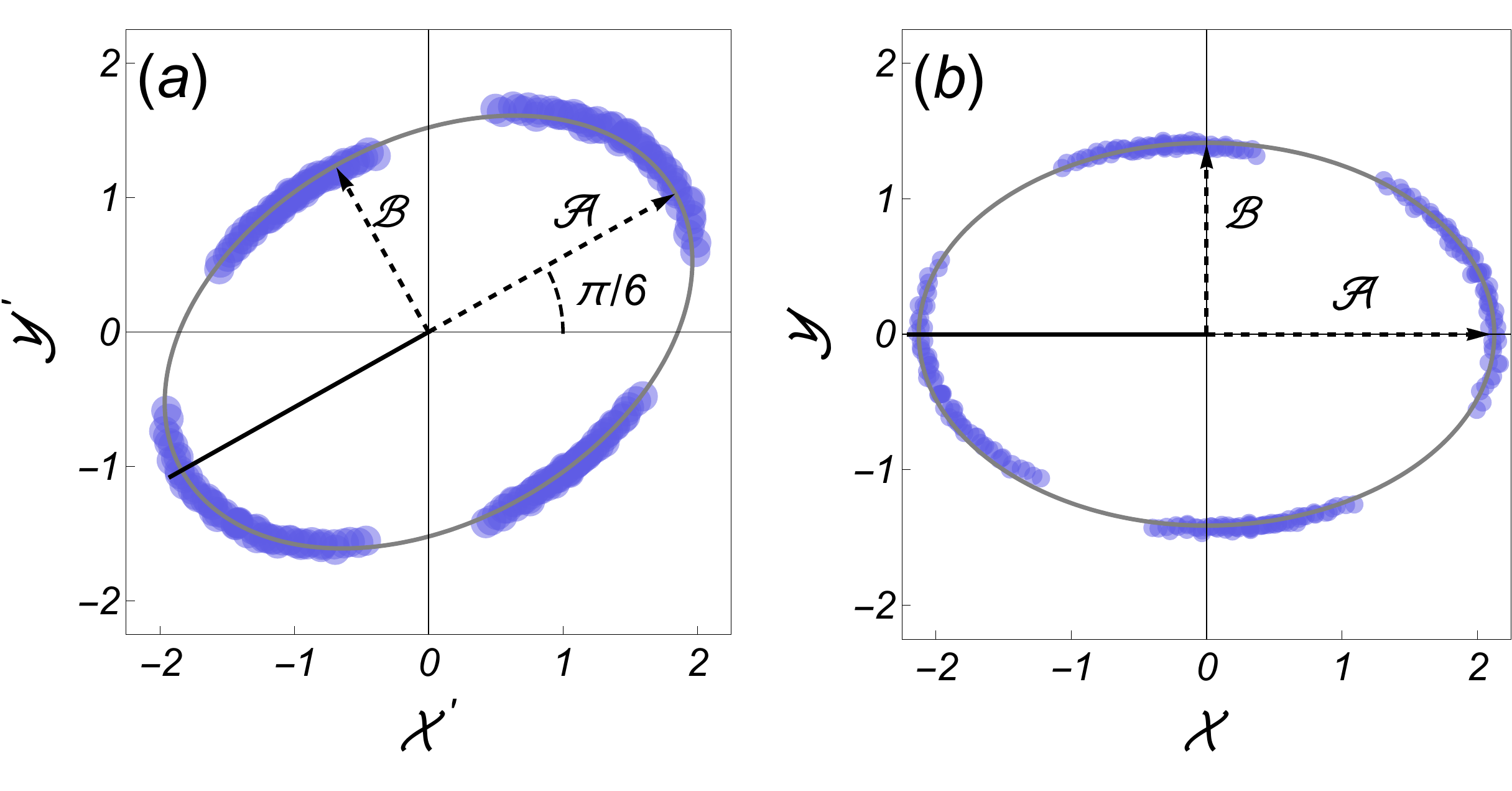}
\caption{ a) Normalized coordinates $\mathcal{X}^\prime$,
$\mathcal{Y}^\prime$ in the standard frame
and b)  normalized coordinates $\mathcal{X}$,
$\mathcal{Y}$ in the frame aligned with the 
singularity's symmetry axes. The symmetry axes form an angle
of $\nu=32.31^{\circ}=0.5640\,\,\mathrm{rad}\approx \pi/6\,\, \mathrm{rad}$
with the $\mathcal{X}^\prime$-axis. In both frames the fit is indicated
with a solid line.}
\label{figure16}
\end{figure}

Using Eqs. (\ref{jxK:gen})-(\ref{jyC:gen}) we calculate
the kinetic and canonical density currents
shown in Fig. \ref{figure14}. Notice that the probability density
evolves by rotating in the counterclockwise direction.
In the same figure we also show the position of the singularity
whose wave function we wish to characterize.

The starting point is a grid of $16\times 16$ nodes centered in the singularity .
In each one of
them the values of the $x$ and $y$ components of the Bohmian velocity are known.
We have intentionally added a $10\%$ error to the velocity components.
A zoom of the Bohmian velocity field of $\psi$ close to the singularity
is exhibited in Fig. \ref{figure15}.

The topological charge is the first parameter that we extract from this map.
Given that the Bohmian stream lines spin in the counterclockwise direction,
$S$ must be positive.
It is readily obtained through Eq. (\ref{topo:kinetic}) by numerically
integrating the Bohmian velocity around a square centered in the
singularity. This procedure yields $S=+3.0012$, which is consistent with
the structure of the wave function in Eq. (\ref{exp:psi}).

The remaining parameters are calculated as follows:
adding the squares of Eqs. (\ref{vBx:app}) and (\ref{vBy:app})
we obtain the equation of an ellipse
\begin{equation}
\left(\frac{\mathcal{X}}{\mathcal{A}}\right)^2+\left(\frac{\mathcal{Y}}{\mathcal{B}}\right)^2=1,
\label{ellipse:one}
\end{equation}
with semi-major and semi-minor axes given by
$\mathcal{A}=\sqrt{\left\vert S\beta/\gamma\right\vert}$,
$\mathcal{B}=\sqrt{\left\vert S\gamma/\beta\right\vert}$ respectively.
The normalized coordinates $\mathcal{X}=X/R=(x-x_j)/R$ and $\mathcal{Y}=Y/R=(y-y_j)/R$
are  in a reference frame oriented along the singularity's symmetry axes and
centered in the singularity.
Bohmian velocities enter the calculation through the normalising length, given by
\begin{equation}
R=\sqrt{\frac{\hbar}{m}}\sqrt[4]{\left(\frac{x-x_j}{v^K_y}\right)^2+\left(\frac{y-y_j}{v^K_x}\right)^2}.
\end{equation}
As the orientation of the singularity is yet unknown we first calculate
$(\mathcal{X}^\prime=X^\prime/R$ and $\mathcal{Y}^\prime=Y^\prime/R$  in the standard frame.
It is worthwhile to notice that the variables $R$ and
consequently $\mathcal{X}^\prime$
and $\mathcal{Y}^\prime$ can be fully computed from the Bohmian velocity map.
Fig. \ref{figure16} (a) shows a plot of the points
$(\mathcal{X}^\prime,\mathcal{Y}^\prime)$ in the
standard frame. From the tilt of the ellipse we calculate the angle formed
by the singularity's symmetry axes and the standard frame. The angle that best aligns the
ellipse is $\nu=32.31^{\circ}=0.5640\,\,\mathrm{rad}\approx \pi/6\,\, \mathrm{rad}$.
After a $32.31^{\circ}$ rotation, the variables in the singularity frame take the form
$\mathcal{X}=\mathcal{X}^\prime \cos\nu+\mathcal{Y}^\prime \sin\nu$ and
$\mathcal{Y}=\mathcal{Y}^\prime \cos\nu-\mathcal{X}^\prime \sin\nu$. The new variables
are plotted in Fig. \ref{figure16} (b).

\begin{figure}
\includegraphics[angle=0,width=0.50 \textwidth]{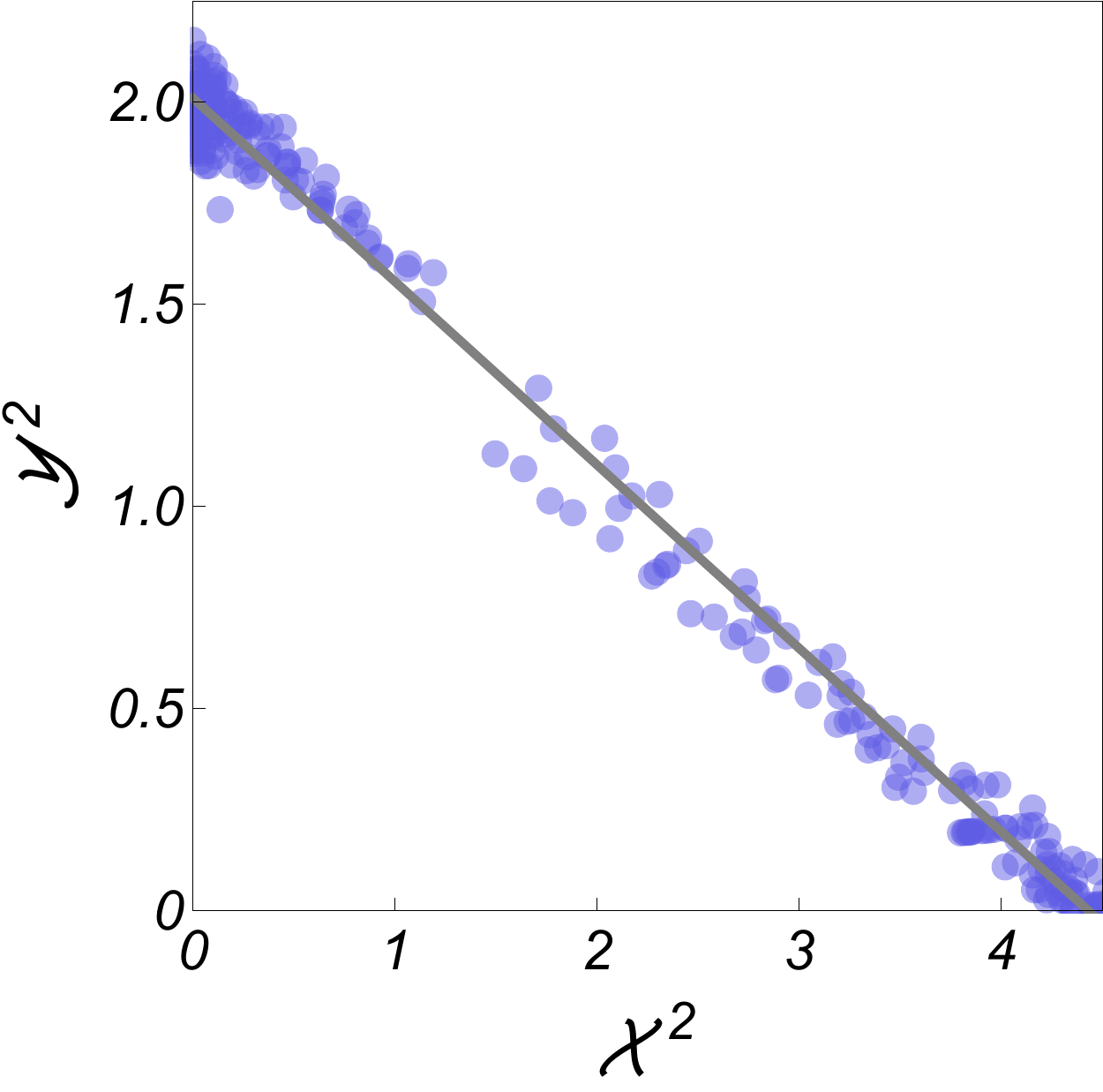}
\caption{ $(X^\prime)^2$ vs. $(Y^\prime)^2$ and fit  $(Y^\prime)^2=-m (X^\prime)^2+c$,
$m=-0.441762$, $c=1.99527$}
\label{figure17}
\end{figure}

By fitting Eq. (\ref{ellipse:one}) (see Fig. \ref{figure17}) we find the semi-major and semi-minor axes
$\mathcal{A}=2.105$ and $\mathcal{B}=1.418$. The fitted curve is shown in
Figs. \ref{figure16} (a) and (b) with a solid line. The topological charge $S$ 
can be newly calculated as $S=\mathcal{A}\mathcal{B}=2.985$.
Finally, the ratio of the $\gamma$ to the $\beta$ parameters
can be determined from 
$\left\vert \beta/\gamma\right\vert=\mathcal{A}/\mathcal{B}=1.485$.

All of these results are consistent with the structure of
the wave function (\ref{exp:psi}).

\section{Conclusions}
\label{conclusions}
We have investigated the relation between phase singularities and the
internal rotational dynamics in electron beams subject to a constant
and uniform magnetic field. To understand different singularity configurations
we have studied superpositions of LG modes having different angular momenta.
In particular we have used HG beams and their positive and negative angular
momentum parts. BEC fields where also examined given their regular singularity
structures.
We have demonstrated 
that the wave function's mathematical form
in the vicinity of a singularity
plays a key role in shaping the velocity profile of the Bohmian
streamlines.
The topological charge, the main parameter defining the singularity's mathematical
structure, can be analytically calculated from the lowest non-vanishing order
of the wave function expanded around the singularity's position.
Conversely, we show  that the shape of the Bohmian stream lines
can be used to estimate the mathematical structure of the singularity.
The method developed here could lead to an experimental procedure
to obtain the relevant parameters of the electron's wave function
starting from a map of the Bohmian velocities.

\acknowledgments
We gratefully appreciate the financial support
of ``Departamento de Ciencias B\'asicas UAM-A" grant numbers
2232214 and 2232215. 
J. C. Sandoval-Santana and V. G. Ibarra-Sierra
would like to acknowledge the support
received from the
``Becas de Posgrado UAM" scholarship numbers
2151800745 and  2112800069.
%We are indebted to Professor J. Grabinsky  
%for the careful reading of the manuscript.

%\bibliography{biblio}
%

\end{document}